\documentclass[sigconf,screen]{acmart}

\renewcommand\footnotetextcopyrightpermission[1]{} 

\usepackage{hyperref}
\usepackage{booktabs}   
\usepackage{subcaption} 
\usepackage[figurename=Fig.]{caption}
\usepackage[tablename=Tab.]{caption}
\usepackage{amsmath}

\usepackage{amssymb}
\usepackage{amsthm}
\usepackage{textcomp}
\usepackage{mathtools}
\usepackage{amscd}
\usepackage{lipsum}
\usepackage{amsfonts}
\usepackage{varwidth}
\usepackage{graphicx}
\usepackage{float}
\usepackage{mathpartir}
\usepackage{multirow}
\usepackage[linesnumbered,ruled,lined,noend]{algorithm2e}
\usepackage[noend]{algorithmic}
\usepackage{mathrsfs}
\usepackage{fancyvrb}
\usepackage{float}
\usepackage{wrapfig}
\usepackage{xspace}
\usepackage[shortlabels]{enumitem}
\usepackage{tkz-base,tikz}
\usepackage{pgfplots}
\usepackage{subcaption}
\usepackage[normalem]{ulem}
\usepackage{listings}
\def\BibTeX{{\rm B\kern-.05em{\sc i\kern-.025em b}\kern-.08em
    T\kern-.1667em\lower.7ex\hbox{E}\kern-.125emX}}

\makeatletter 
\def\arcr{\@arraycr}
\makeatother

\usepackage{cuted}
\usepackage{balance}

\setlist[itemize]{leftmargin=*}
\setlist[enumerate]{leftmargin=*}

\floatstyle{plaintop}
\restylefloat{table}

\theoremstyle{remark}

\makeatletter 
\def\arcr{\@arraycr}
\makeatother

\definecolor{deepblue}{rgb}{0,0,0.5}
\definecolor{deepred}{rgb}{0.6,0,0}
\definecolor{deepgreen}{rgb}{0,0.5,0}
\definecolor{halfgray}{gray}{0.55}
\definecolor{ipythonframe}{RGB}{207, 207, 207}

\definecolor[named]{ACMBlue}{cmyk}{1,0.1,0,0.1}
\definecolor[named]{ACMYellow}{cmyk}{0,0.16,1,0}
\definecolor[named]{ACMOrange}{cmyk}{0,0.42,1,0.01}
\definecolor[named]{ACMRed}{cmyk}{0,0.90,0.86,0}
\definecolor[named]{ACMLightBlue}{cmyk}{0.49,0.01,0,0}
\definecolor[named]{ACMGreen}{cmyk}{0.20,0,1,0.19}
\definecolor[named]{ACMPurple}{cmyk}{0.55,1,0,0.15}
\definecolor[named]{ACMDarkBlue}{cmyk}{1,0.58,0,0.21}

\definecolor{ckeyword}{HTML}{7F0055}
\definecolor{ccomment}{HTML}{3F7F5F}
\definecolor{cnumber}{HTML}{2A0099}
\definecolor{pblue}{rgb}{0.13,0.13,1}
\definecolor{pgreen}{rgb}{0,0.5,0}
\definecolor{pred}{rgb}{0.9,0,0}
\definecolor{pgrey}{rgb}{0.46,0.45,0.48}

\lstdefinestyle{OCaml}{
  language=Caml,
  identifierstyle=\color{black},
  sensitive=true,
  commentstyle=\color{ccomment}\sffamily,
  string=[b]",
  morekeywords={module, struct, open, include, val, nonrec, sig},  
  showstringspaces=false,
  showspaces=false,
  showtabs=false,
  breaklines=true,
  breakatwhitespace=true,
  lineskip=-0.6pt,
  basewidth={0.54em, 0.4em},%
  basicstyle=\scriptsize\ttfamily,
  keywordstyle={\color{ckeyword}\bfseries},
  ndkeywordstyle={\color{ACMDarkBlue}},
  commentstyle={\color{ccomment}\itshape},
  stringstyle={\color{pgreen}},
  numberstyle={\scriptsize\color{cnumber}\ttfamily},
}

\definecolor{eclipseStrings}{RGB}{42,0.0,255}
\definecolor{eclipseKeywords}{RGB}{127,0,85}
\colorlet{numb}{magenta!60!black}

\lstdefinelanguage{json}{
    basicstyle=\normalfont\ttfamily,
    commentstyle=\color{eclipseStrings}, 
    stringstyle=\color{eclipseKeywords}, 
    numbersep=8pt,
    showstringspaces=false,
    breaklines=true,
    frame=single,
    lineskip=-0.6pt,
  	basewidth={0.54em, 0.4em},%
 	basicstyle=\scriptsize\ttfamily,
    string=[s]{"}{"},
    comment=[l]{:\ "},
    morecomment=[l]{:"},
    literate=
        *{0}{{{\color{numb}0}}}{1}
         {1}{{{\color{numb}1}}}{1}
         {2}{{{\color{numb}2}}}{1}
         {3}{{{\color{numb}3}}}{1}
         {4}{{{\color{numb}4}}}{1}
         {5}{{{\color{numb}5}}}{1}
         {6}{{{\color{numb}6}}}{1}
         {7}{{{\color{numb}7}}}{1}
         {8}{{{\color{numb}8}}}{1}
         {9}{{{\color{numb}9}}}{1}
}

\definecolor{shadecolor}{gray}{1.00}
\definecolor{ddarkgray}{gray}{0.75}
\definecolor{darkgray}{gray}{0.30}
\definecolor{light-gray}{gray}{0.87}

\newcommand{\footpatch}{{FootPatch}\xspace}
\newcommand{\saver}{{SAVER}\xspace}
\newcommand{\pulse}{{Pulse}\xspace}
\newcommand{\tool}{{EffFix}\xspace}
\newcommand{\tooluni}{{EffFix\textsubscript{u}}\xspace}
\newcommand{\infertool}{{Infer}\xspace}
\newcommand{\ourinfer}{{Infer-7499c03}\xspace}
\newcommand{\codeql}{{CodeQL}\xspace}

\newcommand{\many}[1]{\overline{#1}}

\newcommand{\eqdef}{\triangleq}

\newcommand{\ite}[3]{\mathsf{{ITE(#1,#2,#3)}}}

\newcommand{\True}{\mathsf{True}}
\newcommand{\False}{\mathsf{False}}
\newcommand{\retz}{\mathsf{{\bf return}}\xspace{}}
\newcommand{\whilez}{\mathsf{{\bf while}}\xspace{}}
\newcommand{\gotoz}{\mathsf{{\bf goto}}\xspace{}}

\newcommand{\constext}[1]{\mathsf{#1}}

\newcommand{\mallocPCword}{\constext{new}}
\newcommand{\freePCword}{\constext{free}}

\newcommand{\malloc}[1]{{\tt{\mallocPCword}}(#1)}
\newcommand{\free}[1]{{\tt{\freePCword}}(#1)}

\newcommand{\asgn}{:=}

\newcommand{\nullc}{{\tt{NULL}}}
\newcommand{\mcode}[1]{{\ensuremath{\tt #1}}}

\newcommand{\domaindetect}{$\mathcal{D}$}
\newcommand{\domainequiv}{$\mathcal{D'}$}
\newcommand{\ok}{\tt{ok}}
\newcommand{\err}{\tt{err}}
\newcommand{\cok}{{\color{deepgreen}\ok}}
\newcommand{\cerr}{{\color{ACMRed}\err}}

\newcommand{\collectfn}[2]{ \mathit{abs}({\color{ACMBlue}#1}, #2)}

\newcommand{\collectrule}[3]{
	\inferrule*{#2}{#3}
}

\newcommand{\pts}{\mapsto}
\newcommand{\pointsto}[2]{#1 \pts #2}
\newcommand{\dealloc}[1]{#1 \not\pts}

\newcommand{\emp}{\mathsf{emp}}
\newcommand{\sep}{\ast}
\newcommand{\isltriple}[3]{[{{\color{ACMBlue} #1}}]~#2~[{\color{ACMBlue}  #3}]}
\newcommand{\metatriple}[6]{(#1,#4,#5,#6)}
\newcommand{\footprint}[4]{(#3,#1:#4)}
\newcommand{\nil}{\mathsf{nil}}

\newcommand{\ret}{\mathsf{ret}}
\newcommand{\allocset}{H}
\newcommand{\deallocset}{D}
\newcommand{\aliasset}{A}

\newcommand{\algo}[1]{\textsc{#1}}
\newtheorem{definition}{Definition}

\begin{document}

\def\sectionautorefname{Sec.}
\def\subsectionautorefname{Sec.}
\def\subsectionautorefname{Sec.}
\def\subsubsectionautorefname{Sec.}
\def\figureautorefname{Fig.}
\def\tableautorefname{Tab.}
\def\equationautorefname{Eq.}

\interfootnotelinepenalty=10000

\newcommand{\mytitle}{Patch Space Exploration using Static Analysis Feedback}

\newcommand{\runningtitle}{\mytitle}

\title{\mytitle}

\author{Yuntong Zhang}
\email{yuntong@comp.nus.edu.sg}
\affiliation{%
  \institution{National University of Singapore\country{Singapore}}
}

\author{Andreea Costea}
\email{andreeac@comp.nus.edu.sg}
\authornote{corresponding author}
\affiliation{%
  \institution{National University of Singapore\country{Singapore}}
}

\author{Ridwan Shariffdeen}
\email{ridwan@comp.nus.edu.sg}
\affiliation{%
  \institution{National University of Singapore\country{Singapore}}
}

\author{Davin McCall}
\email{davin.mccall@oracle.com}
\affiliation{%
  \institution{Oracle Labs\country{Australia}}
}

\author{Abhik Roychoudhury}
\email{abhik@comp.nus.edu.sg}
\affiliation{%
 \institution{National University of Singapore\country{Singapore}}
}

\begin{abstract}
Automated Program Repair (APR) techniques typically rely on a given test-suite to guide the repair process. Apart from the need to provide test oracles, this makes the produced patches prone to test data over-fitting. In this work, instead of relying on test cases, we show how to automatically repair memory safety issues, by leveraging static analysis (specifically Incorrectness Separation Logic) to guide repair.
Our proposed approach learns what a desirable patch is by inspecting how \emph{close}  a patch is to fixing the bug based on the feedback from incorrectness separation logic based static analysis (specifically the Pulse analyser), and turning this information into a distribution of probabilities over context free grammars.  Furthermore, instead of focusing on heuristics for reducing the search space  of patches, we make repair scalable by creating classes of equivalent patches according to the \emph{effect} they have on the symbolic heap, and then invoking the validation oracle  only once per class of patch equivalence. This allows us to efficiently discover repairs even in the presence of a large pool of patch candidates offered by our generic patch synthesis mechanism. Experimental evaluation of our approach was conducted by repairing real world memory errors in OpenSSL, swoole and other subjects. The evaluation results show  the scalability and efficacy of our approach in automatically producing high quality patches.
\end{abstract}

\maketitle
\pagestyle{plain} 

\section{Introduction}
\label{sec:intro}

Despite decades of efforts put into avoiding or mitigating memory safety errors (which are errors in handling memory in native programming languages such as C), recent surveys show that this class of issues still accounts for two of the three most dangerous software weaknesses reported in 2021 \cite{cwe-2021}. 
For example, reports show that  60\% of the high severity security vulnerabilities and millions of user-visible crashes in Android are due to incorrect memory handling, while Google announced that 70\% of all security bugs in Chrome in 2020 are memory safety issues.
Given the ever increasing reliance on and complexity of software, if left unattended, memory safety bugs in legacy code will continue to prevail and would negatively impact the user experience and trust in software. Therefore, providing the tools and technologies to fix such bugs in a timely and efficient manner is a critical endeavour.

\noindent{\bf \emph{Approaches-to-APR}}. 
Advances in automated program repair (APR) techniques \cite{LPR19} show promise
in dealing with the problem of bug repair.
These techniques predominantly use test cases as a specification of program correctness. However, tests are rarely exhaustive, providing only a loose correctness specification, thus making  such techniques  prone to over-fitting to the test data. Furthermore, this conventional generate-and-validate approach assumes the following sequence of steps for each patch candidate: select a patch from a pre-defined search space and validate it for correctness by running the patched program against the given test cases. Repeated for each plausible patch and given a sufficiently large search space, this process turns out to be quite expensive. 

\footpatch \cite{TonderG18} and \saver \cite{HongLLO20}, the state of the art techniques for repairing memory safety bugs, reduce the reliance on test suites for patch validation in favour of using the advances in static analysis to determine the correctness of patches. \footpatch demonstrates that this direction is a promising one, managing to generate fixes for large codebases. \saver further increases the effectiveness of static-analysis based repair by designing a novel representation of the program called object flow graph, which summarizes the program’s
heap-related behavior using static analysis, resulting in a methodology which generates only safe fixes. 

\noindent{\bf \emph{Our-approach-to-APR}}. In this paper, we present a scalable and sound methodology to fixing memory related bugs without the need of test cases or developer intervention. Similar to the state of the art in repairing memory errors, our approach relies on existing sophisticated static analysis tools for finding a semantically rich class of memory bugs. Different from the current state of the art, our approach replaces the conventional patch synthesis followed by test-based validation with a novel synthesis and validation technique which works in tandem towards: efficiently navigating the search space of plausible patches, 
and achieving high repairability with a generic synthesis engine that does not rely on bug-specific fixing strategies.
To achieve this we adapt the advances of Incorrectness Separation Logic (ISL) \cite{Le2022,Raad2020} for precise bug finding 
to the problem of automated program repair. 

In a nutshell, our approach relies on ISL to describe the semantic effect the patch has on the symbolic heap, and to choose correct patches. Since the search space might be quite large, we propose to categorise patches into \emph{equivalence classes} based on their semantic effect, and subsequently only validate one representative patch per class. 
Furthermore, to increase the likelihood of producing mostly correct patches, the synthesis 
checks how ``close'' a patch is to fixing the bug by checking the patch's effect on the bug, and focuses on search spaces which have a high chance of producing plausible patches. 
In particular, we describe the entire space of solutions using a probabilistic context free grammar and learn which of its production rules are most likely to be involved in a plausible patch. 
This allows for a generic, yet efficient synthesis engine, which is not constrained by custom bug templates or specifications. 
\vspace{1em}

The contributions of this work are as follows:
\begin{itemize}
\item a {\em scalable} approach for static analysis driven repair; the approach partitions large search spaces into semantic effect based equivalence classes, enabling \emph{efficient validation} and scalability;

\item a \emph{generic} APR engine based on static analysis which does not require  bug specific templates or specifications to fix a given bug; instead it relies on the feedback from the analyser to understand what a bug and its correct patch are.

\item an \emph{effective} navigation of the solution space based on probabilistic context free grammars, which favours the production rules with higher chance of deriving a plausible patch;
 
\item an \emph{open source} tool, \tool, which implements our approach to fixing memory safety issue. 
\end{itemize}

\section{Motivation and Overview}
We next highlight some of the key aspects of our approach to APR and support these choices by means of examples. 

\noindent \emph{\bf The case for static analysis.}
Consider the null pointer dereference bug (NPE) in \autoref{fig:exnull-manifest}, which is reported in OpenSSL.
Under low memory, \mcode{OPENSSL\_malloc} returns \mcode{NULL}, thus leading to a null pointer dereference during the call to \mcode{memset} which takes \mcode{param} as an argument. The issue here is that explicitly checking \mcode{param} to be a non-null value---as per the fix high-lighted in green---is not a standard practice  within this project since,
  unlike  \mcode{OPENSSL\_malloc}, most \mcode{malloc} wrappers in OpenSSL abort if the result is \mcode{NULL}. The reservations developers have in acknowledging and fixing such bugs is highlighted in the conversations the authors of a static analyser used at Meta had with the OpenSSL maintainers \cite{Le2022}.

\begin{figure}[t]\hspace{1.4em}
{\footnotesize
\begin{subfigure}{0.41\textwidth}
\begin{lstlisting}[language=C,mathescape=true,firstnumber=90,numbers=left,escapeinside={(*@}{@*)}]
VERIFY_PARAM $*$VERIFY_PARAM_new(void){
 VERIFY_PARAM  $*$param; 
 param=OPENSSL_malloc(sizeof(VERIFY_PARAM));
\end{lstlisting}
\vspace{-\baselineskip}
\begin{lstlisting}[language=C,mathescape=true,firstnumber=93,numbers=left,backgroundcolor=\color{green!20}] 
 (+) if (!param)
 (+)        return NULL;
\end{lstlisting}  
\vspace{-\baselineskip}
\begin{lstlisting}[language=C,mathescape=true,firstnumber=95,numbers=left] 
 memset(param, 0, sizeof(VERIFY_PARAM));
 verify_param_zero(param);
 return param; }
  \end{lstlisting}
  \end{subfigure}
  \vspace{-0.5em}
\caption{An NPE bug
  \vspace{-1.5em}
 and its fix in OpenSSL}
\label{fig:exnull-manifest}
}
\end{figure}

To uncover and fix difficult to detect pointer manipulating bugs such as the NPE in  \autoref{fig:exnull-manifest}, the bug detector should understand the semantic 
effect a statement may have on the heap even in exceptional cases. 
This is hardly possible by means of dynamic testing because of the non-deterministic nature of dynamically allocated data structures and the difficulty of tracking alias information, which explains why so many memory related errors in production remain uncovered or unfixed for many years. The shortfalls of testing demonstrates the case for static analysis as a driving engine for automated program repair since static analyses generally quantify over all possible states a program may be in. We leverage the advances in ISL, a logic tailored for proving the presence of memory bugs, to describe the semantic effects programs have on the heap, and to guide the repair process towards the correct patch, i.e. a patch removing the unwanted semantic effects. 

For our running example, an ISL bug detector is able to infer that a call to \mcode{OPENSSL\_malloc} may result in two different valid program states, one corresponding to an empty memory footprint where the allocation fails, and another one where the allocation succeeds with a footprint comprising a single memory cell abstracted by a symbolic variable $X$:

{\small
\noindent$\isltriple{\emp }{\mcode{OPENSSL\_malloc}}{\cok: \pointsto{\text{param}}{\nil}}$ \\
$\isltriple{\emp }{\mcode{OPENSSL\_malloc}}{\cok: \pointsto{\text{param}}{X} * \pointsto{X}{\_}}$ 
}

Informally, the above abstract states (simplified for brevity) read as follows: starting from an ${\color{ACMBlue}\emp}$ty heap, the program may result in a valid state (indicated by the label $\cok$) where the resulting pointer points to $\nil$, or in a valid state where the \mcode{param} points to a symbolic heap location $\text{X}$ which stores an unspecified value $\_$.
The first state causes issues at the call to \mcode{memset} at line 95 (ignoring the fix) which requires \mcode{param} to point to a valid memory location. After the call to  \mcode{memset}, the abstract states change to:

{\small
\noindent$\isltriple{\ok: \pointsto{\text{param}}{\nil}}{\mcode{\qquad ~~\,\,memset}}{\cerr: \pointsto{\text{param}}{\nil}}$ \\
$\isltriple{\ok: \pointsto{\text{param}}{X} * \pointsto{X}{\_}}{\mcode{memset}}{\cok: \pointsto{\text{param}}{\!X} * \pointsto{X}{0}}$ 
}

Since there is no modification in the $\cerr$oneous symbolic state other than the label which changed from $\cok$ to $\cerr$, it seems difficult to automatically derive a fix by simply looking at the program's abstract state. That is why, instead of adopting the abstract-state driven template-based patch search \cite{TonderG18} which would restrict the kind of patches we can derive, we opt for a generic synthesis  based on context free grammars (CFG),
and only use the abstract state for validation. We seek to derive patches that lead to valid abstract states, i.e. no memory safety bugs, while keeping the code's functionality unchanged.

\noindent \emph{\bf The case for equivalence classes.} 
The advantages of a CFG driven synthesis are clear, i.e. genericity and simple machinery, and so are its disadvantages, i.e. poor efficiency due to a large search space which makes validation expensive. 
We aim to keep the advantages of our approach, while striving for efficiency. 
To this purpose, as we gradually derive more patches, we refine the search space into equivalent classes of patches, i.e. patches with \emph{indistinguishable} effects on the symbolic heap, and, by doing so, we need not validate every generated patch but only one representative patch per equivalence class. 

Consider the patches in \autoref{fig:equivpatch}---patches that could be generated for the example in \autoref{fig:exnull-manifest}. Although there are small syntactic differences between them, semantically they are equivalent. This equivalence is made obvious by the representation of the semantic effects these patches have on the symbolic heap depicted below each patch. We simplified the view of the heap, from formulae in ISL to sets of disjoint symbolic memory locations; in particular we use the empty set $\{\}$ to denote an empty memory footprint, the singleton $\{X\}$ to denote a memory footprint comprising a single memory cell, and the implication $\mathsf{param} = \nil \implies  \cok: \{\} \wedge \ret=\nil$ to denote the pair of path condition $\mathsf{param} = \nil$ and corresponding heap abstraction. It becomes evident that all these patches have the same effects on the symbolic heap, and we need only validate one of them to conclude the validity of all the others. The size of one such class may exponentially grow  with the size of the symbolic  heap and the number of existing aliases. 

\begin{figure*}[t]
\centering
{\footnotesize
\begin{subfigure}{0.31\textwidth}
\begin{lstlisting}[language=C,mathescape=true,firstnumber=3,numbers=left]
param = OPENSSL_malloc($\ldots$);
(+) if (!param) 
(+)   return NULL;  
memset(param, 0, $\ldots$);
\end{lstlisting}
\end{subfigure}
\begin{subfigure}{0.31\textwidth}\begin{lstlisting}[language=C,mathescape=true,firstnumber=3,numbers=left]
param = OPENSSL_malloc($\ldots$);
(+) if (!param) 
(+)   return param;  
memset(param, 0, $\ldots$);
\end{lstlisting}
\end{subfigure}
\begin{subfigure}{0.31\textwidth}\begin{lstlisting}[language=C,mathescape=true,firstnumber=3,numbers=left]
param = OPENSSL_malloc($\ldots$);
(+) if (param == NULL) 
(+)   return param;  
memset(param, 0, $\ldots$);
\end{lstlisting}
\end{subfigure}
\begin{subfigure}{0.31\textwidth}
\vspace{1em}
$\mathsf{param} = \nil \implies  \cok: \{\} \wedge \ret=\nil$\\
$\mathsf{param} \neq \nil \implies \cok: \{X\}  \wedge \ret=\mathsf{param}$
\end{subfigure}
\begin{subfigure}{0.31\textwidth}
\vspace{1em}
$\mathsf{param} = \nil \implies \cok: \{\}  \wedge \ret=\mathsf{param}$\\
$\mathsf{param} \neq \nil \implies  \cok: \{X\}  \wedge \ret=\mathsf{param}$
\end{subfigure}
\begin{subfigure}{0.31\textwidth}
\vspace{1em}
$\mathsf{param} = \nil \implies \cok: \{\} \wedge \ret=\mathsf{param}$\\
$\mathsf{param} \neq \nil \implies  \cok: \{X\} \wedge \ret=\mathsf{param}$
\end{subfigure}
}
\vspace{-0.5em}
\caption{Equivalent patches and their effects for the bug in \autoref{fig:exnull-manifest}}\label{fig:equivpatch}
\end{figure*}

\begin{figure*}[t]
\centering
{\footnotesize
\begin{subfigure}{0.31\textwidth}
\begin{lstlisting}[language=C,mathescape=true,firstnumber=3,numbers=left]
param = OPENSSL_malloc($\ldots$);
(+) if (false) 
(+)   return NULL; 
memset(param, 0, $\ldots$);
\end{lstlisting}
\vspace{-1em}
\caption{}\label{fig:nonsol1}
\end{subfigure}
\begin{subfigure}{0.31\textwidth}
\begin{lstlisting}[language=C,mathescape=true,firstnumber=3,numbers=left]
param = OPENSSL_malloc($\ldots$);
(+) if (param!=NULL) 
(+)   return NULL; 
memset(param, 0, $\ldots$);
\end{lstlisting}
\vspace{-1em}
\caption{}\label{fig:nonsol2}
\end{subfigure}
\begin{subfigure}{0.31\textwidth}\begin{lstlisting}[language=C,mathescape=true,firstnumber=3,numbers=left]
param = OPENSSL_malloc($\ldots$);
(+)  param = app_malloc($\ldots$);
memset(param, 0, $\ldots$);
\end{lstlisting}
\vspace{-1em}
\caption{}\label{fig:nonsol4}
\end{subfigure}
}
\vspace{-1em}
\caption{Non-solutions for the bug in \autoref{fig:exnull-manifest}}\label{fig:nonsol}
\vspace{-1em}
\end{figure*}

\noindent \emph{\bf The case for probabilistic context free grammars} (PCFG). 
A large pool of patches generally means that significant time is spent
in generating incorrect patches. Ideally, we would like to spend less time in exploring patches belonging to an equivalence class of incorrect patches, and instead focus on search spaces (in the form of CFG productions) more likely to produce correct patches. 
To do that we equip the CFG with probabilities which indicate the likelihood of a certain 
production rule to be fired in a correct patch. 

Understanding what is a correct patch without a specification is tricky. We break the correctness criterion into four requirements. The first and most obvious requirement is for the patch to actually fix the bug. For example, all patches in figure 
\autoref{fig:equivpatch} fixes the bug, however, the patches in \autoref{fig:nonsol1} and \autoref{fig:nonsol2} are non-solutions since \mcode{NULL} can still flow into \mcode{memset}. We reward the production rules used in generating  the patches in 
\autoref{fig:equivpatch} with a maximum reward, while offering no rewards for those in the incorrect patches since they have no effect on the buggy path. Although an \mcode{if-then} construct is used in both kind of patches, by choosing not to reward it in the incorrect patches, instead of, say, penalizing it, allows us to still explore the space of plausible patches containing \mcode{if-then} but with different condition expressions.

The second requirement is that the patch should introduce no new bugs, and, third, it should only affect the path on which the considered bug manifests, i.e. when  \mcode{param} is \mcode{NULL}.
For example, the patch in  \autoref{fig:nonsol4} is a non-solution:  
although this patch fixes the NPE (\mcode{app\_malloc} is a wrapper which aborts if the allocation is unsuccessful), it also changes the intended functionality of the program since it affects the case where \mcode{param} is not \mcode{NULL} and introduces a potential memory leak. Although a non-solution, we still choose to reward this patch, albeit a very small reward, since it offers an 
insight into how to remove the bug, hence a search space worth exploring further. 

So far we have learnt that \mcode{if-then} is highly likely to be part of a correct patch, and that, although with a lesser probability, \mcode{app\_malloc} can also fix the bug. This setup leads to a correct patch that wraps the \mcode{app\_malloc} in a conditional affecting only the buggy path, which in turn brings us to the fourth concern regarding patch correctness: what should the error handling routine be? Existing solutions for learning the error handling routine \cite{Junhee2022} are bug specific and would restrict the applicability of our approach. 
Instead, we choose to discover as many plausible solutions as allowed by a given time constraint, and give the developer the autonomy of choosing the \emph{preferred class} of correct patches. 

In general, the probabilities ascribed to the CFG are gradually learnt from measuring how close the patch is to fixing a bug and how much the memory footprint changes. For example, the memory footprints of the buggy method in \autoref{fig:exnull-manifest}, the one fixed with a patch from \autoref{fig:equivpatch}, and the one incorrectly fixed with the patch in  \autoref{fig:nonsol4} are, respectively, as follows:

\vspace{0.5em}
{\small
\begin{tabular}{rl}
\autoref{fig:exnull-manifest} & $\mathsf{param} = \nil \implies  \cerr: \{\} \wedge \ret=\nil$\\

{} & $\mathsf{param} \neq \nil \implies \cok: \{X\}  \wedge \ret=\mathsf{param}$
~\\[5pt]

\autoref{fig:equivpatch} & $\mathsf{param} = \nil \implies  \cok: \{\} \wedge \ret=\nil$\\

{}& $\mathsf{param} \neq \nil \implies \cok: \{X\}  \wedge \ret=\mathsf{param}$
~\\[5pt]

\autoref{fig:nonsol4} & $\mathsf{param} = \nil \implies  \text{\tt{abort}}: \{\} $\\

{} & $\mathsf{param} \neq \nil \implies \cerr: \{X,Y\}  \wedge \ret=\mathsf{param}$\\

\end{tabular}
}
~\\

We notice that the memory footprints corresponding to the buggy program and that of the correctly fixed one are almost identical, hence very close,  except for the exit condition label, while the incorrectly patched one differs in the size of the heap too (for the previously non-buggy path) reflecting the calls to the two \mcode{malloc} wrappers as well as the labels for both paths. This justifies why the correct patches receive a high reward.

 \saver \cite{HongLLO20}, the state of the art in repairing memory related bugs, is unable to generate a fix for our running example since the object flow analysis on which it operates manipulates events and non-allocation cannot be modelled as an event. \footpatch does handle null pointer dereferences but its search and template-based methodology cannot always generate fixes on specific paths, if the fix template has not been seen before - leading to restrictive fixes.

\section{Repair Framework}
 
\begin{figure}[t]
\centering
\includegraphics[width=0.47\textwidth]{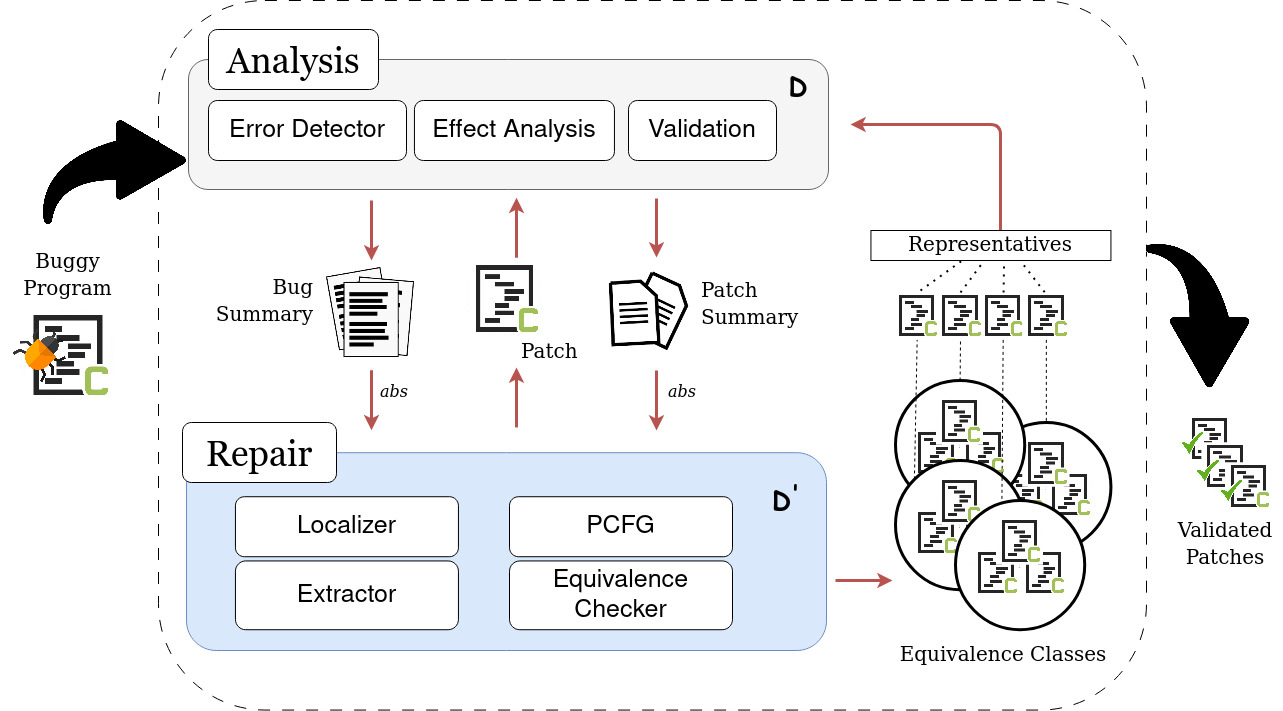}
\caption{Framework Overview}\vspace{-2em}
\label{fig:overview}
\end{figure}

 \autoref{fig:overview} offers a summary view of our APR framework based on static analysis.
  A static bug detector running on an abstract domain \domaindetect~(ISL in our case)
  takes as input a buggy program and creates a summary of the 
   the bug (the footprint of the buggy method, the path condition on which the bugs manifests, and the culprit statement). 
   A patch is then synthesised using a PCFG. We investigate the effects the patch has on the memory footprint by creating a summary of the buggy method after having applied the newly created patch. 
   Synthesised patches are then clustered into equivalence classes according to their effect on the symbolic heap. 
   Before refining the equivalence classes by checking which class this patch belongs to, we further abstract the patch's summary using a simplified abstract domain, so as to avoid having to check for the equivalence of ISL formulae. This meta-domain, \domainequiv, mostly retains information about what memory cells have been allocated and deallocated, and about the program paths and exit conditions. 
   Identifying the patch's equivalence class takes into consideration the ``distance" from the bug, and classifies patches accordingly. 
   If the synthesised patch does not remove the considered bug or if it affects other paths than the buggy ones, then this fact is transmitted to the patch synthesis mechanism  in order for it to fine-tune the probabilities ascribed to the PCFG. In other words, the probabilities implicitly reflect how the search space should be navigated. 
   Finally, each plausible equivalence class is validated by picking only one representative patch per class.

\section{Methodology}

This section describes how bugs are detected, how patches are synthesised using probabilistic context free grammars
and subsequently classified into equivalence classes according to the effect they have on the footprint of the buggy program.

\subsection{Bug Detection}
We build our approach on top of \pulse \cite{pulse-link}, an industry-grade static analysis tool 
which soundly detects memory safety violations. \pulse uses the latest advances in Incorrectness Separation Logic (ISL), a logic tailored to reason about the presence of bugs for heap-manipulating programs. \pulse first abstracts the C input program to an intermediate language, the Smallfoot Intermediate Language (SIL),
and then runs an abstract interpretation engine to check for safety bugs.

\noindent \emph{\bf Program model.} A SIL core set of expressions and commands is depicted in \autoref{fig:lang}. A program in SIL is a sequence of procedures, and a procedure is a composition of heap manipulating commands and standard commands, such as allocation, deallocation, conditionals, etc. The storage model comprises a stack and a heap, where the stack is a function from the set of program and logical variables to values, and the heap is a partial function from symbolic heap locations to values. 
A state thus models a stack and a heap, and together with an environment which tracks the values associated with program and logical variables it models a \pulse world.

\noindent \emph{\bf The abstract domain (\domaindetect).} The abstract domain on which \pulse operates when symbolically executing the SIL commands is depicted in \autoref{fig:logic}: a symbolic heap $\Delta$ comprises a spatial term \mcode{k} and a pure, first order logical formula, $\pi$ to account for pointer aliasing and non-heap information. 
The spatial term 
$\emp$ is an assertion to denote an empty heap,  \mcode{\pointsto{v}{X}}  is the points-to assertion for the program variable \mcode{v}, while 
 \mcode{ \pointsto{Y}{X}} is the points-to for logical variables.  \mcode{ \dealloc{X}} denotes memory deallocation, and  the separation logic conjunction \mcode{  k \sep k}  denotes disjoint sub-heaps. 
 An abstract state $\Phi$ is defined as a pair of a program path $\pi$ and a symbolic heap $\Delta$.
 
\noindent \emph{\bf Bug detection.} 
\pulse uses summaries (specifications) of predefined instructions to infer the summary (specification) of a given piece of code \cite{Le2022}. 
At the core of \pulse is the ISL (under-approximate) triple  $\isltriple{\Phi_\textit{pre}}{c}{\epsilon:\Phi_\mathit{post}}$ which asserts that any final state satisfying $\Phi_\mathit{post}$ is reachable by executing $c$ starting from an initial state satisfying $\Phi_\mathit{pre}$. Furthermore, the exit condition $\epsilon$ indicates either a normal termination, i.e. $\ok$, or a buggy one, i.e. $\err$.
The pair $(\Phi_\textit{pre}, \epsilon:\Phi_\mathit{post})$ describes the effect $c$ has on one program path, 
and a set $F$ (\autoref{fig:logic}) of such effects describes the memory footprint of $c$ where each effect in the set corresponds to a unique program path.

\noindent \emph{\bf Bug description.} A bug report in \pulse comprises the information $\epsilon$ about the bug kind, e.g. null dereference, and the culprit statement  $c$, e.g. the statement which dereferences a null pointer. On top, we record the summary of the method which contains the bug, $F$, and  the path $\pi$ on which the bug manifests. 
A bug is defined in terms of the following tuple:
\\
\centerline{$b ~::=~ \langle \epsilon,  \pi, c, F \rangle$.}

For a bug $b$ we will often refer to an element of the tuple using the dot notation, e.g. $b.F$. The same notation is used throughout the paper for other kinds of tuples as well. 


\begin{figure*}[t]
\[
\;\;\;\;\;\;\;\;\;\;\;
{\small
\begin{array}{l r c l}
 \\[2pt]
 \text{Variables} \qquad v & \multicolumn{3}{l}{
  \text{Alpha-numeric identifiers} \qquad {x, y    }
   \qquad\qquad\
  \text{Locations} \qquad {\textit{loc}    }
  \qquad\quad~
   \text{Pointers} ~~ ptr ~~ ::= ~~ v \mid \nullc}
 \\[10pt]
 \text{Boolean Expression} & b & ::= &
 \True \mid  \False \mid  b ~\vee~ b \mid   b ~\wedge~ b \mid  \neg b \mid x ~\textit{op}_r~ x \mid  \textit{ptr} ~ \textit{op}_p  ~ \textit{ptr}
 \\[2pt]
 \text{Relational Operator} & \textit{op}_r & ::= &  <  ~\mid~ <= ~\mid~ == ~\mid~  !\!\!= ~\mid~ > ~\mid~ >= 
  \qquad\qquad\qquad
   \text{Pointers Operator} ~~~ \textit{op}_p ~~ ::=~~ ==  ~\mid~ !\!\!=
 \\[2pt]
 \text{Heap Manipulation} & s & ::= & v \asgn p \mid  v \asgn [ \textit{ptr} ] \mid  [ \textit{ptr} ] \asgn  \textit{ptr}  \mid   \textit{ptr}  = \malloc{} \mid  \free{v} 
 \\[2pt]
 \text{Commands} &
 c & ::= &
            s 
            \mid x :=  f(\many{x})
            \mid c; c 
            \mid \ite{b}{c}{c}
            \mid \whilez(b) \{c\}
\mid \retz~  \textit{x}  \mid \retz~  \textit{ptr} \mid \gotoz ~ \textit{label}
 \\[10pt]     
 \text{Patch} &
  P & ::= & \text{INSERT} ~c~\textit{loc} ~\mid~   \text{COND} ~\False~\textit{loc}   
\end{array}
}
\]
\vspace{-0.5em}
\caption{Core (Simplified) Programming Language.}
\label{fig:lang}
\end{figure*}


\begin{figure}[t]
\vspace{-0.8em}
 \centering
\[
\!\!\!\!\!\!\!\!\!\!\!\!
{  \small
\begin{array}{l l c l}
 \text{Exit condition} & \epsilon  & ::=& \{\ok,\err, abort\} \\
 \text{Pure term} & \pi  & ::= & b
 \\
 \text{Spatial term} &\mcode{k} & ::= &
 \mcode{\emp \mid \pointsto{v}{X}}
 \mcode{\mid \pointsto{Y}{X}}
 \mcode{\mid \dealloc{X}}
  \mid  \mcode{  k \sep k}
 \\
 \text{Symbolic heap} & \Delta & ::= & \mcode{ k \wedge \pi}
 \\
 \text{State} & \Phi& ::= & \pi; \Delta \mid \exists X.\pi;\Delta
 \\
 \text{Effect} & E & ::= & (\Phi, \epsilon: \Phi)  
 \\
 \text{Footprint} & F & ::= &  \mathit{Set}(E)  
\end{array}
}
\]
\vspace{-0.5em}
\caption{Abstract domain for bug detection (\domaindetect).}
\vspace{-1em}
\label{fig:logic}
\end{figure}

\subsection{Patch Synthesis with Probabilistic CFGs}
In this work, we employ a synthesis mechanism based on \emph{probabilistic context free grammars (PCFG)} tailored for our approach to APR.
In a CFG, a non-terminal symbol may be expanded in \emph{n} different ways,  
e.g. a command $c$ in \autoref{fig:lang} may be expanded in 8 different ways.
In a PCFG, each production rule comes annotated with a probability $p$, denoting the probability of this rule being selected, with the proviso that 
the sum of probabilities of all  \emph{n} production rules should be 1, e.g. 
$\sum_{i=1}^8 p_i =1$ for the production rules of command $c$.

In our approach, instead of annotating the production rules with one probability, 
we do so with a pair of probabilities denoted by  $\langle p^\pi, p^e \rangle$, with the same proviso holding 
separately for each probability in the pair, e.g. $\sum_{i=1}^8 p^\pi_i =1$ and  $\sum_{i=1}^8 p^e_i =1$ for the production rules of command $c$.
This design choice was made so as to be able to navigate the search space of patches from two different dimensions in parallel: 
finding patches with a high-probability of affecting only the path on which the bug was found (corresponding to probability $p^\pi$), 
and finding patches with a high-probability of having an effect on the heap state which fixes the considered bug (corresponding to probability $p^e$). 

Assuming these pair of probabilities are set for each production rule,
(we detail in \autoref{subsec:learnprob} how the probabilities are learnt),
we generate patches by simply traversing the grammar and choosing 
production rules based on the product of $ p^\pi * p^e$, since we consider
the event of generating a patch which affects the buggy path and the event 
of generating a patch with the correct memory effect to be independent of each other.
To avoid the risk of leading to a very large (possibly infinite) parsing tree, we bound the size of the 
tree to a height $h$. 
However, this poses the risk of generating syntactically incorrect patches when 
the height $h$ is reached. To avoid this issue, choosing 
the next production rule is a function of the rule's given probability and height: 
if the height of the generated tree is $h$, we prioritise production rules which lead to syntactically correct patches regardless of their probability; else, the probability is the sole deciding factor in choosing the next production rule.  

\subsection{Patch Clustering}\label{ssec:clustering}

To reduce the cost of patch validation we progressively refine the 
solution space by identifying classes of \emph{equivalent patches}, and proceed
with only validating one representative patch per class. 
Two patches are equivalent if we can show that they lead to patched programs 
which have equivalent memory footprints, or, stated differently, they have the
same effect when applied on the buggy program. 
Given an ISL triple $\isltriple{\Phi_{\textit{pre}}}{\textit{fnc}}{\epsilon:\mathit{\Phi_{\textit{post}}}}$, 
the memory footprint of $\mathit{fnc}$ is described by the two memory snapshots/states, $\mathit{\Phi_{\textit{pre}}}$ and ${\epsilon:\mathit{\Phi_{\textit{post}}}}$, respectively. 
Reasoning about equivalent memory footprints would require
reasoning about equivalent ISL formulas, which in turn requires ISL logic entailment checking. 
To break the dependency of the bug domain and make our approach agnostic to the bug detector, we design a meta abstraction on top of ISL in order to describe a simpler memory snapshot. 
\autoref{fig:equiv} describes the meta domain \domainequiv~
used for equivalence checking, while,
defined as a recursive function $\mathit{abs}$,
\autoref{fig:transform} introduces some of the main abstraction rules for translating a state from ISL to  \domainequiv.
A memory snapshot in \domainequiv~ is described by a tuple $\phi$ comprising
a path $\pi$ in first order logic, 
a set of allocated 
symbolic memory cells \allocset, a set of deallocated 
symbolic memory cells \deallocset, and a set of pointer aliases \aliasset.
An effect is a tuple $e$ which comprises the exit condition $\epsilon$
and two memory snapshots ${\phi_\text{pre}}$ and ${\phi_\text{post}}$, corresponding to the 
inferred precondition and postcondition, respectively.

\begin{figure}[t]
 \centering
\[
\!\!\!\!\!\!\!\!\!\!\!\!\!\!\!\!\!\!\!\!\!\!\!\!\!\!\!\!
{  \small
\begin{array}{l l c l}
 \text{Allocated symbolic heaps} & \multicolumn{3}{l}{\mcode{H} \subseteq \mathit{Loc}}
 \\
 \text{Deallocated symbolic heaps} & \multicolumn{3}{l}{\mcode{D} \subseteq \mathit{Loc}}
  \\
  \text{Aliases} & \multicolumn{3}{l}{\mcode{A} \subseteq	\mathbb{P}(\mathit{Vars} \times \mathit{Vars)}}
 \\
 \text{State} & \phi & ::= & \mcode{\metatriple{\pi}{\epsilon}{\ret}{\allocset}{\deallocset}{\aliasset}}
 \\
 \text{Effect} & e & ::= & \mcode{\footprint{\epsilon}{\ret}{\phi_\text{pre}}{\phi_\text{post}}}
 \\
 \text{Footprint} &\mathcal{F} & ::= & \mathit{Set}(e)
\end{array}
}
\]
\vspace{-0.5em}
\caption{Abstract domain for equivalence checking (\domainequiv).}
\vspace{0.3em}
\label{fig:equiv}
\end{figure}

\begin{figure*}[h]
{\footnotesize
\begin{mathpar}
		\collectrule{Fun}{
		}{
			\collectfn
			{{\pi; \Delta}}{\_} 
			\eqdef
			\collectfn{\Delta}{\metatriple{\pi}{\epsilon}{\ret}{\emptyset}{\emptyset}{\emptyset}}
		}   
	
					\collectrule{Fun}{
													}{\collectfn
													{\mcode{p_1=p_2 }}{\metatriple{\pi}{\epsilon}{\ret}{\allocset}{\deallocset}{\aliasset}} 
													\eqdef
													{\metatriple{\pi}{\epsilon}{\ret}{\allocset}{\deallocset}{\aliasset \cup\{(p_1,p_2)\} }} 
													}    
													\\
        	\collectrule{Fun}{
            {\metatriple{\_}{\epsilon}{\ret}{\allocset'}{\deallocset'}{\aliasset'}} := \collectfn{k}{\metatriple{\pi}{\epsilon}{\ret}{\allocset}{\deallocset}{\aliasset}}
        	\quad
            {\metatriple{\_}{\epsilon}{\ret}{\_}{\_}{\aliasset''}} := \collectfn{\pi'}{\metatriple{\pi}{\epsilon}{\ret}{\allocset}{\deallocset}{\aliasset}}
			}{\collectfn
			{\mcode{k \wedge \pi'}}{\metatriple{\pi}{\epsilon}{\ret}{\allocset}{\deallocset}{\aliasset}} 
			\eqdef
			{\metatriple{\pi}{\epsilon}{\ret}{ \allocset'}{ \deallocset'}{\aliasset' \cup \aliasset''}}
		}   
		\\
	        	\collectrule{Fun}{
	        	{\metatriple{\pi}{\epsilon}{\ret}{\allocset_1}{\deallocset_1}{\aliasset_1}} := \collectfn{k_1}{\metatriple{\pi}{\epsilon}{\ret}{\allocset}{\deallocset}{\aliasset}} 
	        	\quad
	            {\metatriple{\pi}{\epsilon}{\ret}{\allocset_2}{\deallocset_2}{\aliasset_2}} := \collectfn{k_2}{\metatriple{\pi}{\epsilon}{\ret}{\allocset}{\deallocset}{\aliasset}} 
				}{\collectfn
				{\mcode{k_1 \sep k_2}}{\metatriple{\pi}{\epsilon}{\ret}{\allocset}{\deallocset}{\aliasset}} 
				\eqdef
				{\metatriple{\pi}{\epsilon}{\ret}{\allocset_1 \cup \allocset_2}{\deallocset_1 \cup \deallocset_2}{\aliasset_1 \cup \aliasset_2}} 
			}      
			\\
				\collectrule{Fun}{
				}{\collectfn
				{\mcode{ \pointsto{Y}{X}}}{\metatriple{\pi}{\epsilon}{\ret}{\allocset}{\deallocset}{\aliasset}} 
				\eqdef
				{\metatriple{\pi}{\epsilon}{\ret}{\allocset \cup\{Y\}}{\deallocset}{\aliasset}} 
				}    
				
				\collectrule{Fun}{
								}{\collectfn
								{\mcode{\dealloc{X} }}{\metatriple{\pi}{\epsilon}{\ret}{\allocset}{\deallocset}{\aliasset}} 
								\eqdef
								{\metatriple{\pi}{\epsilon}{\ret}{\allocset}{\deallocset \cup \{Y\}}{\aliasset}} 
								}      
\end{mathpar}}
\caption{Abstract domain transformation (\domaindetect $\rightarrow$ \domainequiv).}
\vspace{-1em}
\label{fig:transform}
\end{figure*}

Given the meta domain \domainequiv ~we can now define  indistinguishable (meta-)effects in terms of indistinguishable states.
\begin{definition}[Indistinguishable states]
Two states $\phi_1$ and $\phi_2$ are said to be indistinguishable, denoted by $\phi_1 \approx \phi_2$  if and only if the following condition holds:

\centerline{
$\phi_1.\pi \Leftrightarrow \phi_2.\pi ~\wedge$
$\phi_1.H = \phi_2.H ~\wedge $ 
$\phi_1.D = \phi_2.D.$
}
\noindent where the equality on sets is defined modulo the alias information stored in $\phi_1.A$ and  $\phi_2.A$, respectively. 
\end{definition}

\begin{definition}[Indistinguishable meta-effects]
Two meta effects $e_1$ and $e_2$ are said to be indistinguishable, denoted by $e_1 \approx e_2$,  if and only if the following condition holds:

\centerline{
 $e_1.\epsilon = e_2.\epsilon ~\wedge$
 $e_1.\text{pre} \approx e_2.\text{pre} ~\wedge$
 $e_1.\text{post} \approx e_2.\text{post} $
}
\end{definition}

So far we talked about a memory footprint as if it comprises a single pair of pre- and post-conditions. However, programs are often ascribed multiple such pairs to account
for different behaviours on different program paths. A memory footprint is thus a disjunction of pair of states in ISL, $F$ in \autoref{fig:logic}, which corresponds to a set of effect tuples in the meta-domain \domainequiv, $\mathcal{F}$ in \autoref{fig:equiv}. We define indistinguishable footprints as follows: 

\begin{definition}[Indistinguishable footprints]
Two footprints $\mathcal{F}_1$ and $\mathcal{F}_2$ are said to be indistinguishable, denoted by $\mathcal{F}_1 \approx \mathcal{F}_2$, if and if the following condition holds:

\centerline{$\forall e_1 \in \mathcal{F}_1, \exists e_2 \in \mathcal{F}_2: ~ e_1 \approx e_2$.}
\end{definition}
In other words, two footprints are indistinguishable if they have indistinguishable meta-effects on each path.  
Equivalent patches are now simply defined as:
\begin{definition}[Equivalent patches]
Two patches $P_1$ and $P_2$ which lead to footprints $F_1$ and $F_2$, respectively, 
when applied to the same buggy program, are said to be equivalent  if and only if their corresponding 
footprint meta-abstractions, that is,  $\mathcal{F}_1$ and $\mathcal{F}_2$, respectively, are indistinguishable:
$\mathcal{F}_1 \approx \mathcal{F}_2$.
\end{definition}

We use the above definition of equivalent patches to
progressively partition the search space into classes of equivalent patches. 
The benefit of this partitioning is that we only need to validate one
patch per class of plausible patches.
Given a bug $b$, a class of plausible patches is one where all 
patches $P$ meet the following condition:

\centerline{$\forall e \in P.\mathcal{F} : (e.\text{post}.\pi \Rightarrow b.\pi ) \Rightarrow e.\epsilon = \ok$}
In other words, the path on which the bug manifested 
is now labelled with an $\ok$ exit condition, i.e. the bug is fixed. 

\subsection{Learning Probabilities}\label{subsec:learnprob}

Starting from a PCFG with a uniform distribution (with regards to the pair of 
probabilities), we ascribe probabilities to this PCFG with the aim of
increasing the likelihood of mostly navigating search spaces of plausible patches. 

We adopt a strategy where we reward the production rules which lead to 
a patch that impacts the path the bug manifests on and those which
lead to a patch that affects the bug's memory footprint.  
Let us assume that the PCFG stores weights instead of probabilities
where weight is an integer indicating the total number of rewards a production rule 
 has received---we use a $\textit{token}$ as the measurement unit.
The rewards then carry different weights, depending on how \emph{close} the generated 
patch is to fixing the bug. 
At a high level,  $ p^\pi$ receives a minimal reward, e.g. 1 token, when the patch affects
the path on which the bug manifests but also other paths. 
$ p^\pi$ receives a maximal reward, e.g. 3 tokens, when the patch affects the entire path on which the bug manifests, while the memory footprint on the other paths remains unchanged.
Similarly, $p^e$ is minimally rewarded if the current patch affects the bug's memory footprint without fixing the bug or by fixing the bug and introducing a new one. $p^e$ is maximally rewarded when the patch fixes the bug without introducing new bugs. 
The production rules used to generate a patch are updated according to the above rewarding strategy, before we normalise the weights back into probabilities:
 $p_{c,i} = w_{c,i} / \sum_{j=1}^{n}w_{c,j}$ for the $i^\textit{th}$ production rule of a symbol $c$ with $n$ alternative rules.

\subsection{Patch Location and Ingredients}\label{ssec:location}
\pulse reports the location where the bug manifests, but we would like a fix at its source. For this purpose we adopt and further adapt 
the Spectrum Based Fault Localization or SBFL \cite{sbfl} to static analysis settings.
SBFL requires test suite to generate pass/fail program traces. 
We collect this information from the program's specification which comprises both safe and buggy paths, thus feeding
SBFL with a comprehensive ``test suite" to cover all possible paths discovered by static analysis. 
The patch ingredients such as variables are computed by a simple taint analysis starting from the culprit object. Other ingredients such as constants and labels are collected within the same function scope as the fix location.

\subsection{Putting it all together} 
Now that we have identified most phases of our approach to APR, we
outline how they are interconnected in \autoref{alg:main}. Given a buggy program 
 $\mathcal{P}$, the algorithm populates a map $M$ with classes of plausible patches for the bugs detected by 
 \pulse (line 4). For each bug $b$, it determines the location where the patch should be inserted
 and collects the ingredients for the patch synthesis (lines 6-7). 
 Starting from a uniform distribution 
 of a PCFG $G$ (line 8), the synthesis of each new patch (line 10) triggers
 a refinement of the patch equivalence classes and an update of the probabilities (line 11). Lastly, we validate only the classes of plausible patches (lines 12-13) 
 by choosing a patch representative per class - we use a simple ranking metric which solely takes into account the size of the patch's AST.

\begin{algorithm}[t]
	\caption{\algo{Main}}
	\label{alg:main}
	\textbf{Input}:  a buggy program $\mathcal{P}$\\
	\textbf{Output}: a map M from bugs to sets of patches \\
	M $\leftarrow$ InitMap()\\
   $B$ = detect the bugs in $\mathcal{P}$   \\
   \For{ $b \in B$ }{
   	loc $\leftarrow$  determine the fix location for $b$\\
   $\mathcal{I}$ $\leftarrow$  collect vars and constants in $\mathcal{P}$ related to $b$\\
   G $\leftarrow$  a PCFG with terminals $\mathcal{I}$ and uniform distrib. \\
   $\mathcal{C}$  $\leftarrow$  $\emptyset$ \\
     \While{$P$ = \normalfont{synthesise a patch using} $G, \mathcal{I}$,\,\normalfont{loc}}{
     $\mathcal{C}$,G $\leftarrow$  \textsc{RefineEquivClasses}($\mathcal{C}$,$P$,G,$b$)}
   $\mathcal{C}'$ $\leftarrow$ filter $\mathcal{C}$ for classes of plausible patches;\\
   $\mathcal{C}''$ $\leftarrow$ validate $\mathcal{C'}$ picking one patch per class;\\
   M $\leftarrow$ update M with $b \rightarrow$ rank($\mathcal{C}''$)
    }
\end{algorithm}   

\begin{algorithm}[t]
	\caption{\algo{RefineEquivClasses}}
	\label{alg:equivclasses}
	\textbf{Input}:  a set of existing patch clusters $\mathcal{C}$, a patch $P$, a PCFG $G$, a bug $b$\\
	\textbf{Output}: updated patch clusters $\mathcal{C}$, updated PCFG $G$ \\
   \For{  {cls} $\in$ {$\mathcal{C}$}}{
   \If{$P.\mathcal{F} - b.\mathcal{F} = \normalfont{\text{summary}}(\textit{cls})$}{
   $\mathcal{C}$ $\leftarrow$ add patch $P$ to the class \textit{cls} of $\mathcal{C}$\\
   G $\leftarrow$ update $G$ according to $P$ and $\textit{cls}$
   }}
  \If{$P \notin\,\mathcal{C}$}{$\mathcal{C},~ \textit{cls}$ $\leftarrow$ add $P$ to a new class in $\mathcal{C}$\\
   G $\leftarrow$ update $G$ according to $P$ and $\textit{cls}$
   }
\end{algorithm}

With each new patch 
the equivalence classes are refined as depicted 
in \autoref{alg:equivclasses}. We mentioned in \autoref{ssec:clustering} that two patches are equivalent if their footprints are indistinguishable. Put differently, given a bug $b$, two of its patches are equivalent 
if they affect the buggy program in which $b$ manifests in the same way. To this purpose, we define a distance relation between a patch and a bug as the symmetric set difference between the sets of allocated and deallocated symbolic heaps for each effect in $P$ and its corresponding effect in $b$:

\centerline{$P.\mathcal{F}-b.\mathcal{F} \eqdef \{e_P - e_b | e_P \in P.\mathcal{F} ~\text{and}~ e_b \in b.\mathcal{F} \}$ }
\noindent where $e_P - e_b$, the difference between effects, tracks how the exit condition changed,  $e_P.\epsilon \rightarrow e_b.\epsilon$, the difference between pre-conditions, and 
the difference between post-conditions. 
$P.\mathcal{F}$ and $b.\mathcal{F}$ are the result of recursively applying the abstraction function $\textit{abs}$ over 
$P.{F}$ and $b.{F}$, respectively.
The difference between states is defined as follows:

\centerline{$\phi - \phi_b \eqdef  \{ (\pi, \allocset \ominus \allocset_b, \deallocset \ominus \deallocset_b, \aliasset \cup \aliasset_b)~ |~ \pi \Rightarrow \pi_b ~\text{and}$ }
\qquad\qquad{\raggedright $ \metatriple{\pi}{\epsilon}{\ret}{\allocset}{\deallocset}{\aliasset} = \phi ~\text{and}~ \metatriple{\pi_b}{\epsilon}{\ret}{\allocset_b}{\deallocset_b}{\aliasset_b} = \phi_b$ \}}

It is this distance definition, namely $P.\mathcal{F}-b.\mathcal{F}$, that is used as
class summary to determine patch equivalence (line 4 in \autoref{alg:equivclasses}). 
A benefit of refining the patch equivalence this way is that it allows us to compute the rewards for the PCFG at the equivalence class level, instead of computing them separately for each synthesised patch (line 6 and line 9). 

\setlength{\textfloatsep}{4pt}

\section{Evaluation}

We aim to answer the following research questions about our approach which is embodied in our tool \tool.
\begin{itemize}
    \item RQ1: How does \tool perform against other similar tools?
    \item RQ2: How efficient are the equivalence classes in reducing the validation costs?
    \item RQ3: How effective is the PCFG in guiding the navigation search space of program patches?
\end{itemize}

\textbf{Implementation.} We implemented our approach on top of \pulse\footnote{the version which comes shipped with \ourinfer}, a sound static analyser for bug finding in the \infertool toolchain used at Meta. We use \pulse to detect bugs, to derive method summaries which we then use to inspect the effect patches have on the symbolic heap, and to validate patches. 
We use a number of custom \codeql queries for collecting patch ingredients.
For finding fix locations, we use a bespoke instance of SBFL. 
For checking program path subsumptions we invoke CVC4, and for quantifier elimination when dealing with logical variables in path formulas we use Z3.

\textbf{Dataset.} We constructed our dataset from the benchmarks of \saver \cite{Le2022} and \pulse \cite{HongLLO20} 
collecting all the subjects containing memory leaks from former, and all memory leaks and NPE from OpensSSL, \pulse's benchmark.
 \saver and \footpatch rely on Separation Logic, a logic which over-approximates program states.
 This conservative approach may discover more bugs but it is prone to false positives, thus risking to put APR tools in the position of fixing non-bugs, e.g. fixing a false memory leak may lead to a double free. Instead, we built on \pulse's ISL which under-approximates states, thus missing some bugs, but it guarantees we only fix true bugs.
We only focus on overlapping bugs that all three tools  detect. 
In total, there are 27 memory issues in our benchmark: 20 memory leaks and 7 NPEs.  \autoref{tab:comparison} contains a summary of these bugs.

Before conducting experiments on \tool, we ran \codeql and \pulse checker on each subject to generate static analysis database and bug detection reports, which serve as inputs to \tool.

\subsection{RQ1: Comparison with Other Tools}
\label{sec:rq1}

\begin{table}[t]
\scriptsize
\centering
\renewcommand{\arraystretch}{1.2}
\setlength{\tabcolsep}{2pt}
\caption{Comparison with state-of-the-art memory error repair tools for large scale C programs.}
\label{tab:comparison}
\begin{tabular}{c|c|ccc|ccc}
\hline
\multirow{2}{*}{Subject} & \multirow{2}{*}{\#Bugs} & \multicolumn{3}{c|}{Plausible Patches} & \multicolumn{3}{c}{Correct Patches} \\ 
\cline{3-8}
 &  & \multicolumn{1}{c|}{EffFix} & \multicolumn{1}{c|}{FootPatch} & SAVER & \multicolumn{1}{c|}{EffFix} & \multicolumn{1}{c|}{FootPatch} & SAVER \\ \hline
\multicolumn{8}{c}{Memory Leaks} \\ \hline

Swoole (a4256e4) & 3 & \multicolumn{1}{c|}{2} & \multicolumn{1}{c|}{2} & 2 & \multicolumn{1}{c|}{2} & \multicolumn{1}{c|}{1} & 2 \\ \hline

p11-kit (ead7ara) & 1 & \multicolumn{1}{c|}{1} & \multicolumn{1}{c|}{1} & 0 & \multicolumn{1}{c|}{0} & \multicolumn{1}{c|}{1} & 0 \\ \hline

x264 (d4099dd) & 6 & \multicolumn{1}{c|}{6} & \multicolumn{1}{c|}{0} & 6 & \multicolumn{1}{c|}{4} & \multicolumn{1}{c|}{0} & 3 \\ \hline

Snort-2.9.13 & 8 & \multicolumn{1}{c|}{7} & \multicolumn{1}{c|}{0} & 8 & \multicolumn{1}{c|}{7} & \multicolumn{1}{c|}{0} & 8 \\ \hline

OpenSSL-1.0.1h & 2 & \multicolumn{1}{c|}{2} & \multicolumn{1}{c|}{0} & 0 & \multicolumn{1}{c|}{2} & \multicolumn{1}{c|}{0} & 0 \\ \hline

Total & 20 & \multicolumn{1}{c|}{18} & \multicolumn{1}{c|}{3} & 16 & \multicolumn{1}{c|}{15} & \multicolumn{1}{c|}{2} & 13 \\ \hline

\multicolumn{8}{c}{Null-pointer Dereferences} \\ \hline
OpenSSL-1.0.1h & 5 & \multicolumn{1}{c|}{4} & \multicolumn{1}{c|}{0} & NA & \multicolumn{1}{c|}{2} & \multicolumn{1}{c|}{0} & NA \\ \hline

OpenSSL-3.0.0  & 2 & \multicolumn{1}{c|}{2} & \multicolumn{1}{c|}{0} & NA & \multicolumn{1}{c|}{2} & \multicolumn{1}{c|}{0} & NA \\ \hline

Total & 7 & \multicolumn{1}{c|}{6} & \multicolumn{1}{c|}{0} & 0 & \multicolumn{1}{c|}{4} & \multicolumn{1}{c|}{0} & 0 \\ \hline

\end{tabular}%
\end{table}

We compare the efficacy of \tool against two state-of-the-art APR tools for memory bugs, \saver~\cite{HongLLO20} and \footpatch~\cite{TonderG18}. 
We set a timeout of 20 minutes for \tool and \saver, since most developers prefer APR tools to produce repairs in under 30 minutes \cite{noller2022trust}.
\footpatch was given a timeout of 1 hour because no patch was produced with the 20 minute timeout.

\autoref{tab:comparison} summarizes the results. 
Columns \emph{Plausible Patches} and \emph{Correct Patches} indicate the number of bugs for which each tool found plausible and correct patches, respectively.
A patch is plausible if it passes the analysis check, e.g. \pulse, and correct if it additionally passes manual inspection.
For memory leaks, \tool and \saver are similarly effective, with each tool finding a correct patch for 15 and 13 bugs respectively, while \footpatch finds correct patches for 2 bugs.
For NPEs, \tool finds correct patches for 4 bugs. \saver is not applicable (NA) to NPEs since it uses pre-defined fix strategies. \footpatch applies to NPEs, but did not generate plausible patches.

\begin{figure}[t]
\scriptsize
\centering

\begin{subfigure}[t]{0.47\textwidth}
\centering
\begin{minipage}[t]{0.47\textwidth}
\centering

\includegraphics[width=0.9\textwidth]{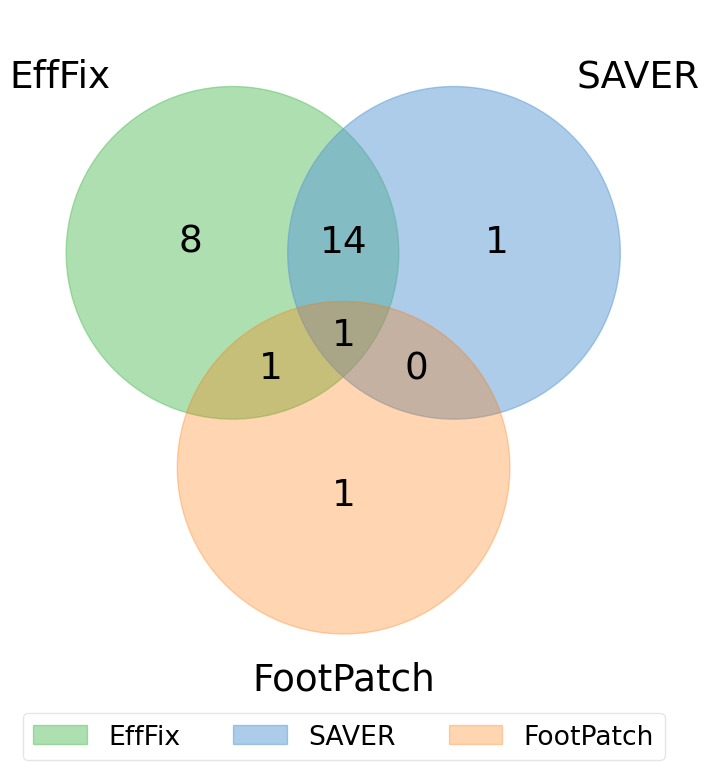}
\caption{Plausible patches}
\label{fig:venn-plausible}
\end{minipage}\hfill
\begin{minipage}[t]{0.47\textwidth}
\centering

\includegraphics[width=0.9\textwidth]{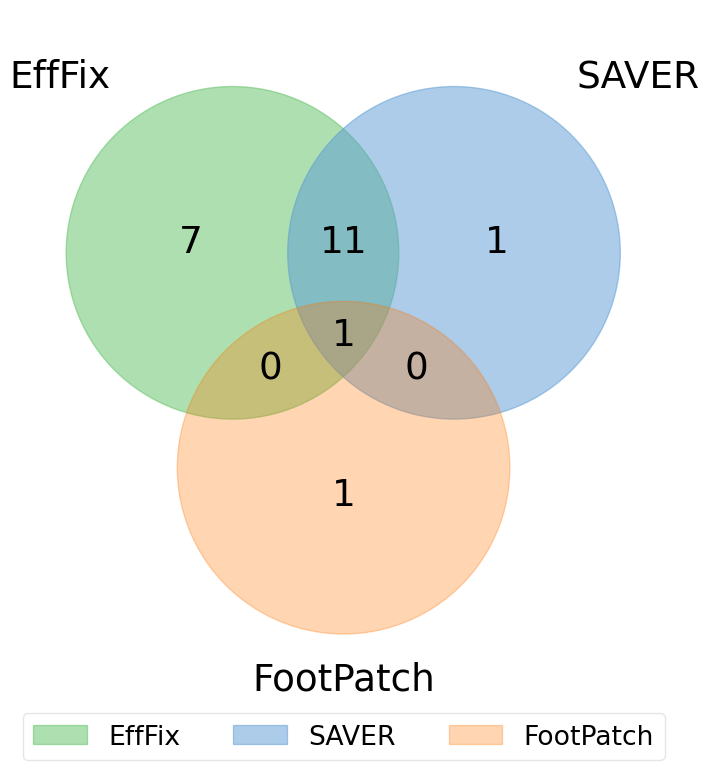}
\caption{Correct patches}
\label{fig:venn-correct}
\end{minipage}

\end{subfigure}

\vspace*{-0.05in}
\caption{Venn diagram of bugs for which each repair tool was able to generate a plausible patch and a correct patch.}
\label{fig:venn-comparison}
\end{figure}

\autoref{fig:venn-plausible} captures the number of \textit{unique} bugs each tool finds plausible patches for. 
\tool found plausible patches for 8 unique bugs while \saver and \footpatch for 1 unique bug each. 
\autoref{fig:venn-correct} shows a similar diagram for correct patches, which shows \tool finds correct patches for 7 unique bugs.

We note that although \tool applies to NPE while \saver does not, \tool still correctly fixes 3 additional memory leaks compared to \saver (out of the 7 unique bugs in \autoref{fig:venn-correct}).
For these 3 bugs, \saver's custom analysis either fails to analyse the bug reported by Infer, or produces a patch with wrong path condition.
\saver generated a correct patch for one bug
for which \tool did not. \tool failed to generate a patch 
due to the large (automatically) constructed search space, which could have been alleviated by using a more strict selection criteria for patch ingredients.

Compared to \tool and \saver, \footpatch found plausible/correct patches for fewer bugs.
One possible reason is that \footpatch searches for candidate repair statements within the program, which could have two consequences. One is that it does not scale well for large codebases such as Snort and OpenSSL.
In fact, \footpatch times out for these two programs in our experiments.
Another consequence is that it fails to find a patch which requires new expressions.

\subsection{RQ2: Efficiency of  Patch Clustering}
\label{sec:rq2}

\begin{table*}[t!]
\scriptsize
\centering
\caption{Results of \tool in fixing memory errors.
Legend: \#Loc: number of fix locations. 
\#PI\textsubscript{p}, \#PI\textsubscript{np}, and \#PI\textsubscript{c} denotes the count for pointer variables, non-pointer variables and constants in patch ingredients, respectively. 
\emph{Space}: estimation of the search space.
\#P\textsubscript{s}: count of syntactically different synthesized patches; 
\#C: number of equivalence classes; 
\#P\textsubscript{lp}: count of locally plausible patches;
\#P\textsubscript{lp-r}: count of {representative} locally plausible patches. 
\#T\textsubscript{p}: count of trials with a plausible patch.
Mean denotes the arithmetic means across bugs;
Count\textsubscript{5} is the count of bugs a plausible patch is generated in all five trials.
}
\label{tab:efffix-results}
\setlength\tabcolsep{4pt}
\renewcommand{\arraystretch}{1.1}
\begin{tabular}{c|c|c|c|c|c|c|c|c|c|c|c|c|c|c|c|c|c}
\hline
\multicolumn{1}{c|}{  \multirow{2}{*}{\textbf{ID}} } & 
\multicolumn{1}{c|}{  \multirow{2}{*}{\textbf{Subject}} } & 
\multicolumn{1}{c|}{  \multirow{2}{*}{\textbf{Type}} } & 
\multicolumn{1}{c|}{  \multirow{2}{*}{\textbf{\#Loc}} } & 
\multicolumn{1}{c|}{  \multirow{2}{*}{\textbf{\#PI\textsubscript{p}}} }&
\multicolumn{1}{c|}{  \multirow{2}{*}{\textbf{\#PI\textsubscript{np}}} }& 
\multicolumn{1}{c|}{  \multirow{2}{*}{\textbf{\#PI\textsubscript{c}}} }& 
\multicolumn{1}{c|}{  \multirow{2}{*}{\textbf{Space}} } &
\multicolumn{5}{c|}{\textbf{\tool}} & 
\multicolumn{5}{c}{\textbf{\tooluni}} \\ 
\cline{9-18}
& & & & & & & & 
\multicolumn{1}{c|}{\textbf{\#P\textsubscript{s}}} & 
\multicolumn{1}{c|}{\textbf{\#C}} & 
\multicolumn{1}{c|}{\textbf{\#P\textsubscript{lp}}} &
\multicolumn{1}{c|}{\textbf{\#P\textsubscript{lp-r}}} &
\multicolumn{1}{c|}{\textbf{\#T\textsubscript{p}}} &

\multicolumn{1}{c|}{\textbf{\#P\textsubscript{s}}} & 
\multicolumn{1}{c|}{\textbf{\#C}} & 
\multicolumn{1}{c|}{\textbf{\#P\textsubscript{lp}}} &
\multicolumn{1}{c|}{\textbf{\#P\textsubscript{lp-r}}} &
\multicolumn{1}{c}{\textbf{\#T\textsubscript{p}}}

\\ \hline\hline
1 & \multirow{3}{*}{Swoole (a4256e4)} & Leak & 2 & 5 & 8 & 6 & 30651250 & 72 & 38 & 0 & 0 & 0 & 67 & 37 & 0 & 0 & 0 \\ 
2 &                                   & Leak & 2 & 2 & 3 & 3 & 445194 & 207 & 43 & 85 & 4.8 & 5  & 212 & 56 & 5.6 & 2.8 & 5 \\
3 &                                   & Leak & 2 & 3 & 1 & 3 & 122474 & 273 & 72 & 74 & 15 & 5 & 325 & 47 & 6.8 & 3.6 & 5  \\ \hline
4 & p11-kit (ead7ara)                 & Leak & 1 & 7 & 2 & 3 & 724341 & 259 & 51 & 31 & 1.8 & 5 & 216 & 58 & 3.6 & 1.0 & 5 \\ \hline
5 & \multirow{6}{*}{x264 (d4099dd)}   & Leak & 1 & 3 & 6 & 3 & 1048837 & 224 & 36 & 37 & 1.4 & 5 & 195 & 38 & 3.6 & 1.0 & 5 \\ 
6 &                                   & Leak & 1 & 3 & 4 & 3 & 492997 & 864 & 239 & 155 & 19 & 5 & 1080 & 144 & 12 & 6.6 & 5 \\ 
7 &                                   & Leak & 1 & 1 & 1 & 3 & 10089 & 575 & 11 & 72 & 1.0 & 5 & 767 & 11 & 11 & 1.0 & 5 \\ 
8 &                                   & Leak & 1 & 3 & 4 & 3 & 492997 & 340 & 120 & 91 & 1.6 & 5 & 408 & 224 & 4.8 & 2.2 & 5 \\ 
9 &                                   & Leak & 1 & 5 & 5 & 3 & 1696881 & 384 & 55 & 97 & 8.4 & 5 & 401 & 46 & 4.0 & 2.8 & 5 \\ 
10 &                                  & Leak & 1 & 6 & 3 & 5 & 2353453 & 360 & 31 & 309 & 13 & 5 & 404 & 106 & 93 & 20 & 5 \\ \hline
11 & \multirow{8}{*}{Snort-2.9.13}    & Leak & 1 & 4 & 3 & 3 & 468777 & 261 & 78 & 30 & 2.6 & 5 & 254 & 107 & 0.2 & 0.2 & 1 \\ 
12 &                                  & Leak & 1 & 4 & 3 & 3 & 468777 & 277 & 33 & 37 & 1.0 & 5 & 273 & 51 & 0.2 & 0.2 & 1 \\ 
13 &                                  & Leak & 2 & 2 & 5 & 3 & 857322 & 224 & 87 & 2.0 & 1.0 & 3 & 220 & 82 & 0.2 & 0.2 & 1 \\ 
14 &                                  & Leak & 2 & 3 & 6 & 3 & 2202554 & 183 & 79 & 3.8 & 0.6 & 3 & 187 & 93 & 0 & 0 & 0 \\ 
15 &                                  & Leak & 2 & 3 & 7 & 4 & 5115674 & 178 & 85 & 0 & 0 & 0 & 190 & 97 & 0 & 0 & 0 \\ 
16 &                                  & Leak & 1 & 6 & 1 & 2 & 136909 & 174 & 67 & 11 & 1.0 & 3 & 192 & 111 & 0.2 & 0.2 & 1 \\ 
17 &                                  & Leak & 1 & 7 & 1 & 3 & 308493 & 181 & 64 & 20 & 1.0 & 5 & 202 & 111 & 0.2 & 0.2 & 1 \\ 
18 &                                  & Leak & 2 & 3 & 2 & 3 & 289034 & 201 & 66 & 9.4 & 1.8 & 3 & 210 & 76 & 0.4 & 0.4 & 1 \\ \hline
19 & \multirow{9}{*}{OpenSSL-1.0.1h}  & Leak & 1 & 1 & 0 & 4 & 153 & 122 & 5.0 & 8.0 & 2.0 & 5 & 127 & 5 & 8.0 & 2.0 & 5 \\ 
20 &                                  & Leak & 2 & 5 & 4 & 5 & 5858482 & 168 & 23 & 58 & 5.6 & 5 & 183 & 33 & 7.4 & 5.0 & 5\\
21 &                                  & Leak & 2 & 9 & 0 & 4 & 304130 & 140 & 24 & 26 & 3.0 & 5 & 143 & 30 & 4.2 & 2.6 & 5\\ 
22 &                                  & Leak & 1 & 3 & 0 & 4 & 6794 & 137 & 14 & 25 & 1.6 & 5 & 132 & 14 & 7.8 & 2.0 & 5\\ 
23 &                                  & NPE & 1 & 1 & 0 & 3 & 153 & 117 & 10 & 8.0 & 1.0 & 5 & 127 & 10 & 8.0 & 1.0 & 5 \\ 
24 &                                  & NPE & - & - & - & - & - & - & - & - & - & - & - & - & - & - & - \\
25 &                                  & NPE & 1 & 1 & 1 & 4 & 19665 & 312 & 14 & 43 & 2.0 & 5 & 324 & 13 & 13 & 2.0 & 5 \\
26 &                                  & NPE & 1 & 2 & 0 & 3 & 981 & 309 & 67 & 14 & 1.8 & 5 & 817 & 95 & 15 & 2.0 & 5 \\
27 &                                  & NPE & 1 & 5 & 1 & 2 & 85089 & 108 & 60 & 23 & 6.0 & 5 & 127 & 95 & 4.6 & 2.8 & 5 \\ \hline
28 & \multirow{2}{*}{OpenSSL-3.0.0}   & NPE & 1 & 1 & 0 & 2 & 153 & 116 & 12 & 8.0 & 2.0 & 5 & 127 & 12 & 8.0 & 2.0 & 5  \\
29 &                                  & NPE & 1 & 1 & 1 & 4 & 22473 & 259 & 21 & 39 & 2.0 & 5 & 320 & 19 & 12 & 2.0 & 5 \\ \hline\hline

\textbf{Mean} &  &  &  & 3.5 & 2.6 & 3.4 & 1935147 & 251 & 54 & \textbf{47} & \textbf{3.6} &  & 294 & 65 & \textbf{8.3} & \textbf{2.4} &  \\
\textbf{Count\textsubscript{5}} &  &  &  &  &  &  &  &  & & & & \textbf{23}  & & & &  & \textbf{19}  \\ \hline

\end{tabular}
\end{table*}

We evaluated \tool's strategy of clustering patches based on their effects.
\autoref{tab:efffix-results} details our results\footnote{\noindent We target two extra OpenSSL bugs which could not be included in the previous experiment for fairness reasons since \footpatch cannot detect them.}. We focus on the columns under \emph{\tool}, and postpone the discussion of those under  \emph{\tooluni}
to \autoref{sec:rq3}.
To counter for the randomness in the patch synthesis component,
we conducted the experiments for five trials and report the average results where appropriate.
We used a 20-minute timeout for each run, which includes ingredients collection, patch synthesis and clustering.
After the timeout, all patches in plausible clusters are considered as \textit{locally plausible}  (column~\#P\textsubscript{lp}) - the bug was fixed locally in the function.
Since all patches within one cluster are equivalent in the defined abstract domain, only 
one representative patch per cluster is selected as 
candidate for validation (based on its AST size).
We refer to these patches as the \textit{representative} locally plausible patches (column~\#P\textsubscript{lp-r}).
A plausible patch is found when a representative locally plausible patch passes the
whole-program validation.

\textbf{Results.} 
Column \#P\textsubscript{lp} and \#P\textsubscript{lp-r} highlight the effect of patch clustering.
On average, \tool generated 47 locally plausible patches for each bug, and, courtesy to patch clustering, only an average of 3.6 patches are selected for validation purposes. 
In other words, patch clustering reduced the validation efforts by $\sim$13x
with the validation oracle being invoked 3.6 times on average for each bug instead of 47 times.
The reduction in validation costs benefits not only the automated validation oracles such as static analyzers, but also the human developers who might want to examine the plausible patches. 
We note that \tool did not generate plausible patches for 3 bugs (Bug 1, 15 and 24) in all trials. 
The main reason for not finding plausible patches within the timeout is likely the resulted large search space (Bug 1 and 15). 
This larger search space is due to the relatively higher numbers of fix locations and other patch ingredients.
Besides, \tool failed to generate plausible patches for Bug 24 because its bug trace spans multiple functions, which is not supported by our prototype.

\subsection{RQ3: Effectiveness of Probabilistic Grammar}\label{sec:rq3}

\begin{figure}[t]
\scriptsize
\centering

\includegraphics[width=\columnwidth]{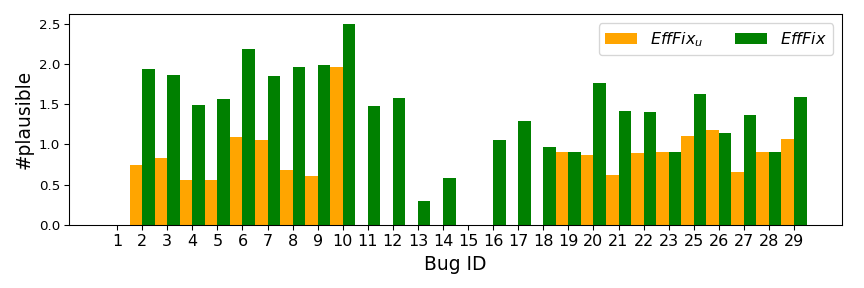}
\caption{Distribution of the average plausible patch count (in log scale) for each bug, for \tool and \tooluni.}
\label{fig:plausible-distribution}
\end{figure}

We next investigate the effects of using a probabilistic grammar to navigate the search space. 
We performed an ablation study by disabling the probability learning in the PCFG.
In other words, the same PCFG with a uniform probability distribution 
is used for both  the patch synthesis and the clustering process.
We refer to this version of \tool as \tooluni (with uniform probability distribution).

\textbf{Results.} The results of evaluating \tooluni are shown in \autoref{tab:efffix-results}, under the columns for \emph{\tooluni}.
Overall, \tooluni generated plausible patches for less number of bugs (19 vs. 23) when considering all five trials, as compared to \tool.
In general, \tooluni failed to generate plausible patches for bugs whose patch requires specific path conditions and pointers (e.g. \texttt{if (a == -1) free(p1);}).
These patches can be sparse in the search space, and a random exploration is less likely to reach the region of plausible patches.
In contrast, \tool  progressively reaches this region by identifying patches in \textit{nearby} regions and biasing the search towards that neighbourhood.
For example, \tool would consider the patch \texttt{if (true) free(p1);} as affecting the bug's memory footprint, and consider the patch \texttt{if (a <= -1) free(p2);} as affecting the bug's path.
By rewarding the production rules used to derive these two patches, the probabilistic grammar can bias the search towards the neighbourhood of plausible patches.

Moreover, the results also show that \tooluni finds lesser locally plausible patches on average, compared to \tool (8.3 vs 47). 
The difference in numbers of plausible patches (that passed whole-program validation)
is then captured in Figure~\ref{fig:plausible-distribution}, which shows the numbers of plausible patches for each bug in log scale.
For 4 bugs with small search spaces (Bug 19, 23, 26, 28) 
the results are similar.
However, overall \tool generated significantly more plausible patches than \tooluni.
This difference is likely due to the search bias: if the search is guided towards regions of plausible patches, more plausible patches would be synthesized within the same time budget.
Higher number of plausible patches also results in a higher number of plausible \textit{regions} being explored: on average, \tool finds 3.6 plausible clusters while \tooluni finds 2.4.
Nonetheless, \tooluni synthesized more patches on average (294 vs. 251)
and created more clusters (65 vs. 54) indicating that it could potentially explore more different regions in the search space.
We note that although it synthesized less patches and explored fewer regions, \tool focused on more plausible regions, thus being more effective.

\subsubsection*{Impact of Probabilities}

\begin{figure*}[h]
\scriptsize
\centering
    \begin{subfigure}[]{0.12\textwidth}
        \centering
       \includegraphics[width=0.95\textwidth]{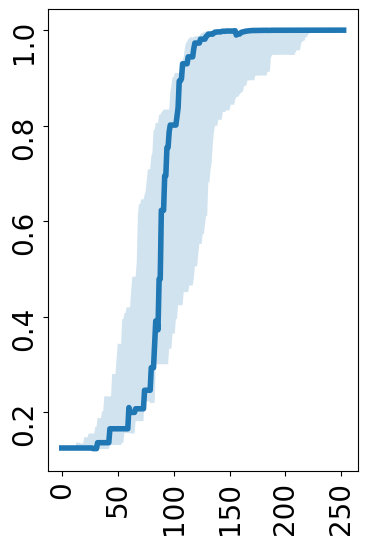}
       \label{fig:snort-1}
       \caption{Bug 11}
    \end{subfigure}%
    ~
    \begin{subfigure}[]{0.12\textwidth}
        \centering
       \includegraphics[width=0.95\textwidth]{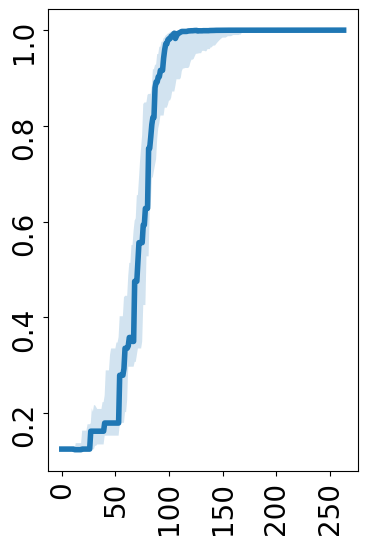}
       \label{fig:snort-2}
       \caption{Bug 12}
    \end{subfigure}%
    ~
    \begin{subfigure}[]{0.12\textwidth}
        \centering
       \includegraphics[width=0.95\textwidth]{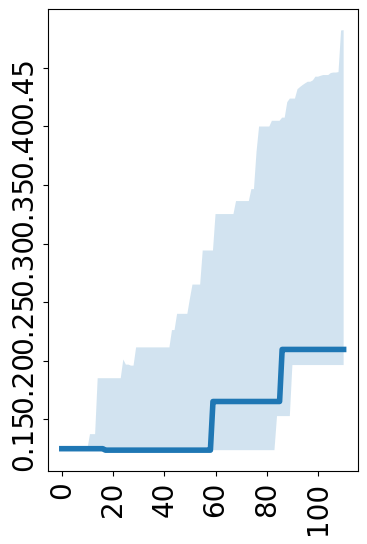}
       \label{fig:snort-3}
       \caption{Bug 13}
    \end{subfigure}%
    ~
    \begin{subfigure}[]{0.12\textwidth}
        \centering
       \includegraphics[width=0.95\textwidth]{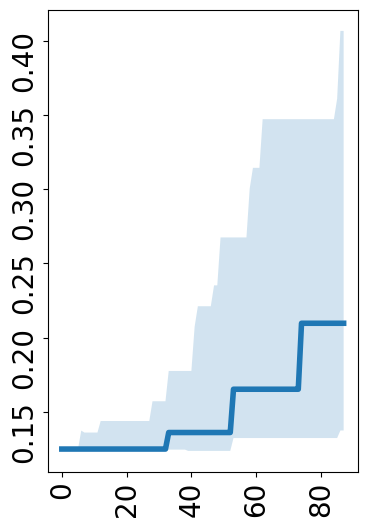}
       \label{fig:snort-4}
       \caption{Bug 14}
    \end{subfigure}%
    ~
        \begin{subfigure}[]{0.12\textwidth}
        \centering
       \includegraphics[width=0.95\textwidth]{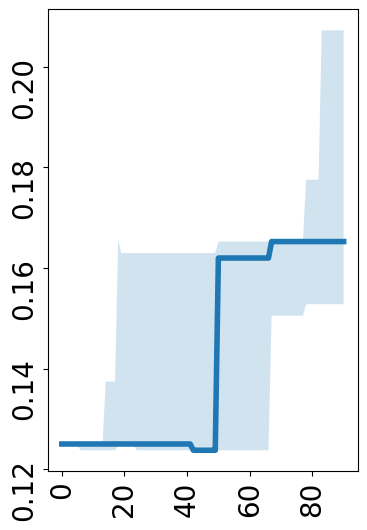}
       \label{fig:snort-5}
       \caption{Bug 15}
    \end{subfigure}%
    ~
    \begin{subfigure}[]{0.12\textwidth}
        \centering
       \includegraphics[width=0.95\textwidth]{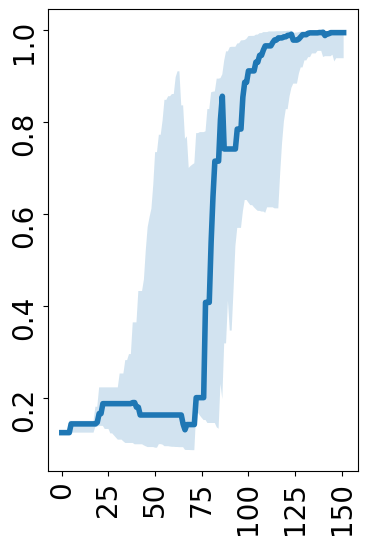}
       \label{fig:snort-6}
       \caption{Bug 16}
    \end{subfigure}%
    ~
    \begin{subfigure}[]{0.12\textwidth}
        \centering
       \includegraphics[width=0.95\textwidth]{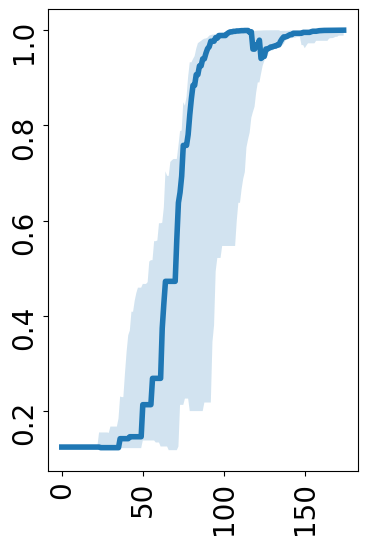}
       \label{fig:snort-7}
       \caption{Bug 17}
    \end{subfigure}%
    ~
    \begin{subfigure}[]{0.12\textwidth}
        \centering
       \includegraphics[width=0.95\textwidth]{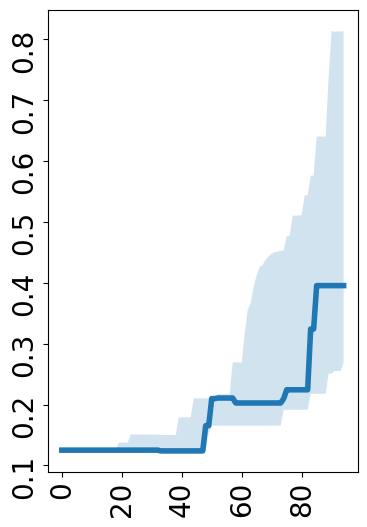}
       \label{fig:snort-8}
       \caption{Bug 18}
    \end{subfigure}%

\caption{Median distribution of the probability change for a correct region for each bug in subject Snort-2.9.13. }
\label{fig:probability-distribution}
\vspace{-2em}
\end{figure*}

We next share some insights on how the probabilities are being learnt to guide the search. We conducted this study on all bugs in the Snort subject, since \tool has more varying performance among bugs (in terms of \#T\textsubscript{p}) within this subject.
All these eight bugs require a fix on a specific path.
We examine whether the search has been gradually directed towards the neighbourhood of conditional patches in each \tool run.
The probability of entering this region at any timestamp is computed by considering all paths in the PCFG derivation tree leading to this sentential form at that timestamp.
Figure~\ref{fig:probability-distribution} plots the probabilities for entering this region over time, for Bug 11-18.
The solid line shows the median trend across 5 trials, and the shaded region denotes 1st-3rd quantile (25\%-75\%).
The maximum x-axis values vary for each plot, because a different number of fix locations were being considered for each bug, and, for brevity, we only plot one location per bug.
The plots show that, with PCFG, the probability of entering the correct region increased over time, directing the search towards this region for all eight bugs.
Another observation is that, although all increasing, the trends for each bug are increasing at different rates.
In fact, this difference is aligned with the stats in \autoref{tab:efffix-results} (column \#P\textsubscript{lp} under \tool). 
For bugs that reached a high probability early and stayed at high probability for some time (e.g Bug 11, 12, 17), \tool was finding considerably more plausible patches, compared to bugs that have a slower increasing trend.
Overall, the plots demonstrate that PCFG can gradually guide the search towards the plausible regions in the search space.

\subsection{Discussion}
\textbf{Limitations.} \tool fixes memory errors but may change the code's intended functionality.
A human oracle currently checks for plausible but incorrect patches which break functionality beyond changes in the heap's shape. 
To ensure that we do not fix false positive bugs, we chose to build on top of \pulse since it has been shown to be sound, although incomplete. This takes care of one of the concerns with static analysis - where it is known to produce lots of false positives.

\textbf{Threats to Validity.} The benchmarks we chose for evaluation might not be representative for the classes of bugs we tackle, but we had to restrict ourselves to the bugs discovered by \pulse. Also, we evaluate \tool against \saver and \footpatch.
However, these tools each tackle different categories of bugs, e.g. \saver cannot handle NPE but fixes use-after-free and double-free, while \footpatch targets resource leaks too. 

\section{Related Work}
{\bf Automated Program Repair.}
Many program repair techniques have been studied in the last decade, largely  for the purpose of fixing logical-errors. Recently it has recieved attention in fixing security vulnerabilities~\cite{senx,extractfix}, race conditions~\cite{hippodrome}, intelligent tutoring~\cite{clara} etc. Program repair techniques can be classified into semantic repair~\cite{angelix,CPR}, search based repair~\cite{genprog,par} or learning based repair~\cite{prophet, sequencer}. Search based repair techinques are known as generate and validate techniques, which heuristically search for a candidate patch in a space of program edits and validate to find a correct patch. Generally, validation is done using dynamic analysis with the aid of a test-suite.  \tool  uses static analysis to validate the generated patches. Using a logic based semantic reasoning, \tool provides additional evidence of correctness for the generated patches, thereby avoiding the patch over-fitting problem~\cite{Qi15,Smith15} as well. 
Fixing memory errors has been studied previously using dynamic analysis~\cite{extractfix,senx,durieux17}, static analysis~\cite{Qing15,Junhee18,TonderG18,HongLLO20,Junhee2022} and combination of both~\cite{Hua16}. Dynamic approaches require a running test case as a witness for the memory error, and have been shown to be effective in fixing buffer-overflows~\cite{extractfix,senx}, null pointer dereference errors~\cite{durieux17}. Our work is closely related to static analysis based repair of memory errors~\cite{TonderG18,HongLLO20}. FootPatch~\cite{TonderG18} generates patches for heap property violations detected using Infer~\cite{infer}. Similarly, SAVER~\cite{HongLLO20} generates safe patches for memory errors detected by Infer~\cite{infer} and was shown to be scalable for larger programs. In both techniques, the patch generated is directly tied to the class of error reported by the static analyser. In contrast, \tool uses a generalized grammar to synthesise patches of arbirary types. Using a probabilistic grammar \tool can dynamically adjust the probabilities to guide the search to correctly identify repair templates which leads to finding more plausible patches.

{\bf Equivalence Classes.} Equivalence relations have been shown to benefit many search problems involving large search spaces such as mutation testing~\cite{rene14,Ma16,Bo17} and compiler testing~\cite{vu14,Sun16}. Recently, it was demonstrated to be effective for APR as well~\cite{MechtaevGTR18}. Equivalence relations can be used to explore larger patch spaces more efficiently. Value based test-equivalence used in~\cite{MechtaevGTR18}, partitions the patch space based on runtime values observed during test executions. In contrast, \tool defines a equivalence relation based on effect analysis.

{\bf Probabilistic Grammar.} Augmenting probabilities with grammar production rules has been shown to be useful in program synthesis~\cite{bhaisaheb23,Ruyi20} and software fuzzing~\cite{Eberlein20,Soremekun22}. Using a probabilistic grammar a software fuzzer can generate inputs based on production rule prioritization. In particular, previous work~\cite{Soremekun22} has shown that evolving a probabilistic grammar can direct the search towards interesting inputs by favouring specific production rules. In contrast, \tool uses a probabilistic grammar to generate program edits rather than program inputs. It evolves the probabilities to find a plausible patch by prioritizing the most promising production rules. 

\section{Conclusion}
\label{sec:conclusion}
We presented an automated program repair approach for fixing memory safety issues guided by static analysis. In our workflow, static analysis is used to both discover and fix a bug, thus alleviating the classic over-fitting issue that test-based approaches normally suffer from. The approach is generic in that it does not require patch templates or bug specifications. To cope with the large search space such a generic method entails, we propose (i) effective patch space navigation by means of probabilistic context-free grammars or PCFGs,  and (ii) efficient patch validation by clustering patches into equivalence classes according to the \emph{effect} they have on the symbolic heap. 

The effect analysis on patches can be potentially extended to other use cases in the future, such as learning the effects of existing error handling routines in the program, which could help program repair/synthesis techniques to generate semantically correct error handling code.

\balance
\bibliographystyle{ACM-Reference-Format}
\bibliography{references}


\begin{thebibliography}{37}


\ifx \showCODEN    \undefined \def \showCODEN     #1{\unskip}     \fi
\ifx \showDOI      \undefined \def \showDOI       #1{#1}\fi
\ifx \showISBNx    \undefined \def \showISBNx     #1{\unskip}     \fi
\ifx \showISBNxiii \undefined \def \showISBNxiii  #1{\unskip}     \fi
\ifx \showISSN     \undefined \def \showISSN      #1{\unskip}     \fi
\ifx \showLCCN     \undefined \def \showLCCN      #1{\unskip}     \fi
\ifx \shownote     \undefined \def \shownote      #1{#1}          \fi
\ifx \showarticletitle \undefined \def \showarticletitle #1{#1}   \fi
\ifx \showURL      \undefined \def \showURL       {\relax}        \fi
\providecommand\bibfield[2]{#2}
\providecommand\bibinfo[2]{#2}
\providecommand\natexlab[1]{#1}
\providecommand\showeprint[2][]{arXiv:#2}

\bibitem[cwe(2022)]%
        {cwe-2021}
 \bibinfo{year}{2022}\natexlab{}.
\newblock \bibinfo{title}{The 2021 Common Weakness Enumeration Top 25 Most
  Dangerous Software Weaknesses}.
\newblock
\newblock
\newblock
\shownote{\url{https://cwe.mitre.org/top25/archive/2021/2021_cwe_top25.html}}.


\bibitem[Abreu et~al\mbox{.}(2007)]%
        {sbfl}
\bibfield{author}{\bibinfo{person}{Rui Abreu}, \bibinfo{person}{Peter
  Zoeteweij}, {and} \bibinfo{person}{Arjan JC~Van Gemund}.}
  \bibinfo{year}{2007}\natexlab{}.
\newblock \showarticletitle{On the accuracy of spectrum-based fault
  localization}. In \bibinfo{booktitle}{\emph{Testing: Academic and Industrial
  Conference Practice and Research Techniques - {MUTATION}
  ({TAICPART-MUTATION})}}.
\newblock


\bibitem[Bhaisaheb et~al\mbox{.}(2023)]%
        {bhaisaheb23}
\bibfield{author}{\bibinfo{person}{Shabbirhussain Bhaisaheb},
  \bibinfo{person}{Shubham Paliwal}, \bibinfo{person}{Rajaswa Patil},
  \bibinfo{person}{Manasi Patwardhan}, \bibinfo{person}{Lovekesh Vig}, {and}
  \bibinfo{person}{Gautam Shroff}.} \bibinfo{year}{2023}\natexlab{}.
\newblock \showarticletitle{Program Synthesis for Complex {QA} on Charts via
  Probabilistic Grammar Based Filtered Iterative Back-Translation}. In
  \bibinfo{booktitle}{\emph{Findings of the Association for Computational
  Linguistics: EACL 2023}}. \bibinfo{publisher}{Association for Computational
  Linguistics}, \bibinfo{address}{Dubrovnik, Croatia},
  \bibinfo{pages}{2456--2470}.
\newblock
\urldef\tempurl%
\url{https://aclanthology.org/2023.findings-eacl.189}
\showURL{%
\tempurl}


\bibitem[Calcagno and Distefano(2011)]%
        {infer}
\bibfield{author}{\bibinfo{person}{Cristiano Calcagno} {and}
  \bibinfo{person}{Dino Distefano}.} \bibinfo{year}{2011}\natexlab{}.
\newblock \showarticletitle{Infer: An Automatic Program Verifier for Memory
  Safety of C Programs}. In \bibinfo{booktitle}{\emph{Proceedings of the Third
  International Conference on NASA Formal Methods}} (Pasadena, CA)
  \emph{(\bibinfo{series}{NFM'11})}. \bibinfo{publisher}{Springer-Verlag},
  \bibinfo{address}{Berlin, Heidelberg}, \bibinfo{pages}{459–465}.
\newblock
\showISBNx{9783642203978}


\bibitem[Chen et~al\mbox{.}(2021)]%
        {sequencer}
\bibfield{author}{\bibinfo{person}{Zimin Chen}, \bibinfo{person}{Steve
  Kommrusch}, \bibinfo{person}{Michele Tufano}, \bibinfo{person}{Louis-Noël
  Pouchet}, \bibinfo{person}{Denys Poshyvanyk}, {and} \bibinfo{person}{Martin
  Monperrus}.} \bibinfo{year}{2021}\natexlab{}.
\newblock \showarticletitle{SequenceR: Sequence-to-Sequence Learning for
  End-to-End Program Repair}.
\newblock \bibinfo{journal}{\emph{IEEE Transactions on Software Engineering}}
  \bibinfo{volume}{47}, \bibinfo{number}{9} (\bibinfo{year}{2021}),
  \bibinfo{pages}{1943--1959}.
\newblock
\urldef\tempurl%
\url{https://doi.org/10.1109/TSE.2019.2940179}
\showDOI{\tempurl}


\bibitem[Contractor and Rivero(2022)]%
        {clara}
\bibfield{author}{\bibinfo{person}{Maheen~Riaz Contractor} {and}
  \bibinfo{person}{Carlos~R. Rivero}.} \bibinfo{year}{2022}\natexlab{}.
\newblock \showarticletitle{Improving Program Matching To Automatically Repair
  Introductory Programs}. In \bibinfo{booktitle}{\emph{Intelligent Tutoring
  Systems: 18th International Conference, ITS 2022, Bucharest, Romania, June 29
  – July 1, 2022, Proceedings}} (Bucharest, Romania).
  \bibinfo{publisher}{Springer-Verlag}, \bibinfo{address}{Berlin, Heidelberg},
  \bibinfo{pages}{323–335}.
\newblock
\showISBNx{978-3-031-09679-2}
\urldef\tempurl%
\url{https://doi.org/10.1007/978-3-031-09680-8_30}
\showDOI{\tempurl}


\bibitem[Costea et~al\mbox{.}(2023)]%
        {hippodrome}
\bibfield{author}{\bibinfo{person}{Andreea Costea}, \bibinfo{person}{Abhishek
  Tiwari}, \bibinfo{person}{Sigmund Chianasta}, \bibinfo{person}{Kishore R},
  \bibinfo{person}{Abhik Roychoudhury}, {and} \bibinfo{person}{Ilya Sergey}.}
  \bibinfo{year}{2023}\natexlab{}.
\newblock \showarticletitle{Hippodrome: Data Race Repair Using Static Analysis
  Summaries}.
\newblock \bibinfo{journal}{\emph{ACM Trans. Softw. Eng. Methodol.}}
  \bibinfo{volume}{32}, \bibinfo{number}{2}, Article \bibinfo{articleno}{41}
  (\bibinfo{date}{mar} \bibinfo{year}{2023}), \bibinfo{numpages}{33}~pages.
\newblock
\showISSN{1049-331X}
\urldef\tempurl%
\url{https://doi.org/10.1145/3546942}
\showDOI{\tempurl}


\bibitem[Durieux et~al\mbox{.}(2017)]%
        {durieux17}
\bibfield{author}{\bibinfo{person}{Thomas Durieux}, \bibinfo{person}{Benoit
  Cornu}, \bibinfo{person}{Lionel Seinturier}, {and} \bibinfo{person}{Martin
  Monperrus}.} \bibinfo{year}{2017}\natexlab{}.
\newblock \showarticletitle{Dynamic patch generation for null pointer
  exceptions using metaprogramming}. In \bibinfo{booktitle}{\emph{2017 IEEE
  24th International Conference on Software Analysis, Evolution and
  Reengineering (SANER)}}. \bibinfo{pages}{349--358}.
\newblock
\urldef\tempurl%
\url{https://doi.org/10.1109/SANER.2017.7884635}
\showDOI{\tempurl}


\bibitem[Eberlein et~al\mbox{.}(2020)]%
        {Eberlein20}
\bibfield{author}{\bibinfo{person}{Martin Eberlein}, \bibinfo{person}{Yannic
  Noller}, \bibinfo{person}{Thomas Vogel}, {and} \bibinfo{person}{Lars
  Grunske}.} \bibinfo{year}{2020}\natexlab{}.
\newblock \showarticletitle{Evolutionary Grammar-Based Fuzzing}. In
  \bibinfo{booktitle}{\emph{Search-Based Software Engineering}},
  \bibfield{editor}{\bibinfo{person}{Aldeida Aleti} {and}
  \bibinfo{person}{Annibale Panichella}} (Eds.). \bibinfo{publisher}{Springer
  International Publishing}, \bibinfo{address}{Cham},
  \bibinfo{pages}{105--120}.
\newblock
\showISBNx{978-3-030-59762-7}


\bibitem[Gao et~al\mbox{.}(2015)]%
        {Qing15}
\bibfield{author}{\bibinfo{person}{Qing Gao}, \bibinfo{person}{Yingfei Xiong},
  \bibinfo{person}{Yaqing Mi}, \bibinfo{person}{Lu Zhang},
  \bibinfo{person}{Weikun Yang}, \bibinfo{person}{Zhaoping Zhou},
  \bibinfo{person}{Bing Xie}, {and} \bibinfo{person}{Hong Mei}.}
  \bibinfo{year}{2015}\natexlab{}.
\newblock \showarticletitle{Safe Memory-Leak Fixing for C Programs}. In
  \bibinfo{booktitle}{\emph{Proceedings of the 37th International Conference on
  Software Engineering - Volume 1}} (Florence, Italy)
  \emph{(\bibinfo{series}{ICSE '15})}. \bibinfo{publisher}{IEEE Press},
  \bibinfo{pages}{459–470}.
\newblock
\showISBNx{9781479919345}


\bibitem[Gao et~al\mbox{.}(2021)]%
        {extractfix}
\bibfield{author}{\bibinfo{person}{Xiang Gao}, \bibinfo{person}{Bo Wang},
  \bibinfo{person}{Gregory~J. Duck}, \bibinfo{person}{Ruyi Ji},
  \bibinfo{person}{Yingfei Xiong}, {and} \bibinfo{person}{Abhik Roychoudhury}.}
  \bibinfo{year}{2021}\natexlab{}.
\newblock \showarticletitle{Beyond Tests: Program Vulnerability Repair via
  Crash Constraint Extraction}.
\newblock \bibinfo{journal}{\emph{ACM Trans. Softw. Eng. Methodol.}}
  \bibinfo{volume}{30}, \bibinfo{number}{2}, Article \bibinfo{articleno}{14}
  (\bibinfo{date}{Feb.} \bibinfo{year}{2021}), \bibinfo{numpages}{27}~pages.
\newblock
\showISSN{1049-331X}
\urldef\tempurl%
\url{https://doi.org/10.1145/3418461}
\showDOI{\tempurl}


\bibitem[Hong et~al\mbox{.}(2020)]%
        {HongLLO20}
\bibfield{author}{\bibinfo{person}{Seongjoon Hong}, \bibinfo{person}{Junhee
  Lee}, \bibinfo{person}{Jeongsoo Lee}, {and} \bibinfo{person}{Hakjoo Oh}.}
  \bibinfo{year}{2020}\natexlab{}.
\newblock \showarticletitle{{SAVER:} scalable, precise, and safe memory-error
  repair}. In \bibinfo{booktitle}{\emph{ICSE}}. \bibinfo{publisher}{{ACM}},
  \bibinfo{pages}{271--283}.
\newblock
\urldef\tempurl%
\url{https://doi.org/10.1145/3377811.3380323}
\showDOI{\tempurl}


\bibitem[{Huang} et~al\mbox{.}(2019)]%
        {senx}
\bibfield{author}{\bibinfo{person}{Z. {Huang}}, \bibinfo{person}{D. {Lie}},
  \bibinfo{person}{G. {Tan}}, {and} \bibinfo{person}{T. {Jaeger}}.}
  \bibinfo{year}{2019}\natexlab{}.
\newblock \showarticletitle{Using Safety Properties to Generate Vulnerability
  Patches}. In \bibinfo{booktitle}{\emph{2019 IEEE Symposium on Security and
  Privacy (SP)}}. \bibinfo{pages}{539--554}.
\newblock
\showISSN{2375-1207}
\urldef\tempurl%
\url{https://doi.org/10.1109/SP.2019.00071}
\showDOI{\tempurl}


\bibitem[Ji et~al\mbox{.}(2020)]%
        {Ruyi20}
\bibfield{author}{\bibinfo{person}{Ruyi Ji}, \bibinfo{person}{Jingjing Liang},
  \bibinfo{person}{Yingfei Xiong}, \bibinfo{person}{Lu Zhang}, {and}
  \bibinfo{person}{Zhenjiang Hu}.} \bibinfo{year}{2020}\natexlab{}.
\newblock \showarticletitle{Question Selection for Interactive Program
  Synthesis}. In \bibinfo{booktitle}{\emph{Proceedings of the 41st ACM SIGPLAN
  Conference on Programming Language Design and Implementation}} (London, UK)
  \emph{(\bibinfo{series}{PLDI 2020})}. \bibinfo{publisher}{Association for
  Computing Machinery}, \bibinfo{address}{New York, NY, USA},
  \bibinfo{pages}{1143–1158}.
\newblock
\showISBNx{9781450376136}
\urldef\tempurl%
\url{https://doi.org/10.1145/3385412.3386025}
\showDOI{\tempurl}


\bibitem[Just et~al\mbox{.}(2014)]%
        {rene14}
\bibfield{author}{\bibinfo{person}{Ren\'{e} Just}, \bibinfo{person}{Michael~D.
  Ernst}, {and} \bibinfo{person}{Gordon Fraser}.}
  \bibinfo{year}{2014}\natexlab{}.
\newblock \showarticletitle{Efficient Mutation Analysis by Propagating and
  Partitioning Infected Execution States}. In
  \bibinfo{booktitle}{\emph{Proceedings of the 2014 International Symposium on
  Software Testing and Analysis}} (San Jose, CA, USA)
  \emph{(\bibinfo{series}{ISSTA 2014})}. \bibinfo{publisher}{Association for
  Computing Machinery}, \bibinfo{address}{New York, NY, USA},
  \bibinfo{pages}{315–326}.
\newblock
\showISBNx{9781450326452}
\urldef\tempurl%
\url{https://doi.org/10.1145/2610384.2610388}
\showDOI{\tempurl}


\bibitem[Kim et~al\mbox{.}(2013)]%
        {par}
\bibfield{author}{\bibinfo{person}{Dongsun Kim}, \bibinfo{person}{Jaechang
  Nam}, \bibinfo{person}{Jaewoo Song}, {and} \bibinfo{person}{Sunghun Kim}.}
  \bibinfo{year}{2013}\natexlab{}.
\newblock \showarticletitle{Automatic patch generation learned from
  human-written patches}. In \bibinfo{booktitle}{\emph{2013 35th International
  Conference on Software Engineering (ICSE)}}. \bibinfo{pages}{802--811}.
\newblock
\urldef\tempurl%
\url{https://doi.org/10.1109/ICSE.2013.6606626}
\showDOI{\tempurl}


\bibitem[Le et~al\mbox{.}(2022)]%
        {Le2022}
\bibfield{author}{\bibinfo{person}{Quang~Loc Le}, \bibinfo{person}{Azalea
  Raad}, \bibinfo{person}{Jules Villard}, \bibinfo{person}{Josh Berdine},
  \bibinfo{person}{Derek Dreyer}, {and} \bibinfo{person}{Peter~W. O'Hearn}.}
  \bibinfo{year}{2022}\natexlab{}.
\newblock \showarticletitle{Finding Real Bugs in Big Programs with
  Incorrectness Logic}.
\newblock \bibinfo{journal}{\emph{Proc. ACM Program. Lang.}}
  \bibinfo{volume}{6}, \bibinfo{number}{OOPSLA1}, Article
  \bibinfo{articleno}{81} (\bibinfo{date}{apr} \bibinfo{year}{2022}),
  \bibinfo{numpages}{27}~pages.
\newblock
\urldef\tempurl%
\url{https://doi.org/10.1145/3527325}
\showDOI{\tempurl}


\bibitem[Le et~al\mbox{.}(2014)]%
        {vu14}
\bibfield{author}{\bibinfo{person}{Vu Le}, \bibinfo{person}{Mehrdad Afshari},
  {and} \bibinfo{person}{Zhendong Su}.} \bibinfo{year}{2014}\natexlab{}.
\newblock \showarticletitle{Compiler Validation via Equivalence modulo Inputs}.
  In \bibinfo{booktitle}{\emph{Proceedings of the 35th ACM SIGPLAN Conference
  on Programming Language Design and Implementation}} (Edinburgh, United
  Kingdom) \emph{(\bibinfo{series}{PLDI '14})}. \bibinfo{publisher}{Association
  for Computing Machinery}, \bibinfo{address}{New York, NY, USA},
  \bibinfo{pages}{216–226}.
\newblock
\showISBNx{9781450327848}
\urldef\tempurl%
\url{https://doi.org/10.1145/2594291.2594334}
\showDOI{\tempurl}


\bibitem[{Le Goues} et~al\mbox{.}(2012)]%
        {genprog}
\bibfield{author}{\bibinfo{person}{C. {Le Goues}}, \bibinfo{person}{T.
  {Nguyen}}, \bibinfo{person}{S. {Forrest}}, {and} \bibinfo{person}{W.
  {Weimer}}.} \bibinfo{year}{2012}\natexlab{}.
\newblock \showarticletitle{GenProg: A Generic Method for Automatic Software
  Repair}.
\newblock \bibinfo{journal}{\emph{IEEE Transactions on Software Engineering}}
  \bibinfo{volume}{38}, \bibinfo{number}{1} (\bibinfo{date}{Jan}
  \bibinfo{year}{2012}), \bibinfo{pages}{54--72}.
\newblock
\showISSN{1939-3520}
\urldef\tempurl%
\url{https://doi.org/10.1109/TSE.2011.104}
\showDOI{\tempurl}


\bibitem[Le~Goues et~al\mbox{.}(2019)]%
        {LPR19}
\bibfield{author}{\bibinfo{person}{Claire Le~Goues}, \bibinfo{person}{Michael
  Pradel}, {and} \bibinfo{person}{Abhik Roychoudhury}.}
  \bibinfo{year}{2019}\natexlab{}.
\newblock \showarticletitle{Automated Program Repair}.
\newblock \bibinfo{journal}{\emph{Commun. ACM}}  \bibinfo{volume}{62}
  (\bibinfo{year}{2019}).
\newblock
Issue 12.


\bibitem[Lee et~al\mbox{.}(2018)]%
        {Junhee18}
\bibfield{author}{\bibinfo{person}{Junhee Lee}, \bibinfo{person}{Seongjoon
  Hong}, {and} \bibinfo{person}{Hakjoo Oh}.} \bibinfo{year}{2018}\natexlab{}.
\newblock \showarticletitle{MemFix: Static Analysis-Based Repair of Memory
  Deallocation Errors for C}. In \bibinfo{booktitle}{\emph{Proceedings of the
  2018 26th ACM Joint Meeting on European Software Engineering Conference and
  Symposium on the Foundations of Software Engineering}} (Lake Buena Vista, FL,
  USA) \emph{(\bibinfo{series}{ESEC/FSE 2018})}.
  \bibinfo{publisher}{Association for Computing Machinery},
  \bibinfo{address}{New York, NY, USA}, \bibinfo{pages}{95–106}.
\newblock
\showISBNx{9781450355735}
\urldef\tempurl%
\url{https://doi.org/10.1145/3236024.3236079}
\showDOI{\tempurl}


\bibitem[Lee et~al\mbox{.}(2022)]%
        {Junhee2022}
\bibfield{author}{\bibinfo{person}{Junhee Lee}, \bibinfo{person}{Seongjoon
  Hong}, {and} \bibinfo{person}{Hakjoo Oh}.} \bibinfo{year}{2022}\natexlab{}.
\newblock \showarticletitle{NPEX: Repairing Java Null Pointer Exceptions
  without Tests}. In \bibinfo{booktitle}{\emph{2022 IEEE/ACM 44th International
  Conference on Software Engineering (ICSE)}}. \bibinfo{pages}{1532--1544}.
\newblock
\urldef\tempurl%
\url{https://doi.org/10.1145/3510003.3510186}
\showDOI{\tempurl}


\bibitem[Long and Rinard(2016)]%
        {prophet}
\bibfield{author}{\bibinfo{person}{Fan Long} {and} \bibinfo{person}{Martin
  Rinard}.} \bibinfo{year}{2016}\natexlab{}.
\newblock \showarticletitle{Automatic Patch Generation by Learning Correct
  Code}. In \bibinfo{booktitle}{\emph{Proceedings of the 43rd Annual ACM
  SIGPLAN-SIGACT Symposium on Principles of Programming Languages}} (St.
  Petersburg, FL, USA) \emph{(\bibinfo{series}{POPL '16})}.
  \bibinfo{publisher}{Association for Computing Machinery},
  \bibinfo{address}{New York, NY, USA}, \bibinfo{pages}{298--312}.
\newblock
\showISBNx{9781450335492}
\urldef\tempurl%
\url{https://doi.org/10.1145/2837614.2837617}
\showDOI{\tempurl}


\bibitem[Ma and Kim(2016)]%
        {Ma16}
\bibfield{author}{\bibinfo{person}{Yu-Seung Ma} {and}
  \bibinfo{person}{Sang-Woon Kim}.} \bibinfo{year}{2016}\natexlab{}.
\newblock \showarticletitle{Mutation Testing Cost Reduction by Clustering
  Overlapped Mutants}.
\newblock \bibinfo{journal}{\emph{J. Syst. Softw.}} \bibinfo{volume}{115},
  \bibinfo{number}{C} (\bibinfo{date}{may} \bibinfo{year}{2016}),
  \bibinfo{pages}{18–30}.
\newblock
\showISSN{0164-1212}
\urldef\tempurl%
\url{https://doi.org/10.1016/j.jss.2016.01.007}
\showDOI{\tempurl}


\bibitem[Mechtaev et~al\mbox{.}(2018)]%
        {MechtaevGTR18}
\bibfield{author}{\bibinfo{person}{Sergey Mechtaev}, \bibinfo{person}{Xiang
  Gao}, \bibinfo{person}{Shin~Hwei Tan}, {and} \bibinfo{person}{Abhik
  Roychoudhury}.} \bibinfo{year}{2018}\natexlab{}.
\newblock \showarticletitle{Test-Equivalence Analysis for Automatic Patch
  Generation}.
\newblock \bibinfo{journal}{\emph{TOSEM}} \bibinfo{volume}{27},
  \bibinfo{number}{4} (\bibinfo{year}{2018}), \bibinfo{pages}{15:1--15:37}.
\newblock
\urldef\tempurl%
\url{https://doi.org/10.1145/3241980}
\showDOI{\tempurl}


\bibitem[Mechtaev et~al\mbox{.}(2016)]%
        {angelix}
\bibfield{author}{\bibinfo{person}{Sergey Mechtaev}, \bibinfo{person}{Jooyong
  Yi}, {and} \bibinfo{person}{Abhik Roychoudhury}.}
  \bibinfo{year}{2016}\natexlab{}.
\newblock \showarticletitle{Angelix: Scalable Multiline Program Patch Synthesis
  via Symbolic Analysis}. In \bibinfo{booktitle}{\emph{Proceedings of the 38th
  International Conference on Software Engineering}} (Austin, Texas)
  \emph{(\bibinfo{series}{ICSE '16})}. \bibinfo{publisher}{Association for
  Computing Machinery}, \bibinfo{address}{New York, NY, USA},
  \bibinfo{pages}{691–701}.
\newblock
\showISBNx{9781450339001}
\urldef\tempurl%
\url{https://doi.org/10.1145/2884781.2884807}
\showDOI{\tempurl}


\bibitem[Noller et~al\mbox{.}(2022)]%
        {noller2022trust}
\bibfield{author}{\bibinfo{person}{Yannic Noller}, \bibinfo{person}{Ridwan
  Shariffdeen}, \bibinfo{person}{Xiang Gao}, {and} \bibinfo{person}{Abhik
  Roychoudhury}.} \bibinfo{year}{2022}\natexlab{}.
\newblock \showarticletitle{Trust enhancement issues in program repair}. In
  \bibinfo{booktitle}{\emph{Proceedings of the 44th International Conference on
  Software Engineering}}. \bibinfo{pages}{2228--2240}.
\newblock


\bibitem[O'Hearn et~al\mbox{.}({[n.\,d.]})]%
        {pulse-link}
\bibfield{author}{\bibinfo{person}{Peter O'Hearn} {et~al\mbox{.}}}
  \bibinfo{year}{[n.\,d.]}\natexlab{}.
\newblock \bibinfo{title}{{Infer:Pulse}}.
\newblock \bibinfo{howpublished}{\url{https://fbinfer.com/docs/checker-pulse}}.
\newblock


\bibitem[Qi et~al\mbox{.}(2015)]%
        {Qi15}
\bibfield{author}{\bibinfo{person}{Zichao Qi}, \bibinfo{person}{Fan Long},
  \bibinfo{person}{Sara Achour}, {and} \bibinfo{person}{Martin Rinard}.}
  \bibinfo{year}{2015}\natexlab{}.
\newblock \showarticletitle{An Analysis of Patch Plausibility and Correctness
  for Generate-and-Validate Patch Generation Systems}. In
  \bibinfo{booktitle}{\emph{Proceedings of the 2015 International Symposium on
  Software Testing and Analysis}} (Baltimore, MD, USA)
  \emph{(\bibinfo{series}{ISSTA 2015})}. \bibinfo{publisher}{Association for
  Computing Machinery}, \bibinfo{address}{New York, NY, USA},
  \bibinfo{pages}{24–36}.
\newblock
\showISBNx{9781450336208}
\urldef\tempurl%
\url{https://doi.org/10.1145/2771783.2771791}
\showDOI{\tempurl}


\bibitem[Raad et~al\mbox{.}(2020)]%
        {Raad2020}
\bibfield{author}{\bibinfo{person}{Azalea Raad}, \bibinfo{person}{Josh
  Berdine}, \bibinfo{person}{Hoang-Hai Dang}, \bibinfo{person}{Derek Dreyer},
  \bibinfo{person}{Peter O'Hearn}, {and} \bibinfo{person}{Jules Villard}.}
  \bibinfo{year}{2020}\natexlab{}.
\newblock \showarticletitle{Local Reasoning About the Presence of Bugs:
  Incorrectness Separation Logic}. In \bibinfo{booktitle}{\emph{Computer Aided
  Verification}}, \bibfield{editor}{\bibinfo{person}{Shuvendu~K. Lahiri} {and}
  \bibinfo{person}{Chao Wang}} (Eds.). \bibinfo{publisher}{Springer
  International Publishing}, \bibinfo{address}{Cham},
  \bibinfo{pages}{225--252}.
\newblock
\showISBNx{978-3-030-53291-8}


\bibitem[Shariffdeen et~al\mbox{.}(2021)]%
        {CPR}
\bibfield{author}{\bibinfo{person}{Ridwan Shariffdeen}, \bibinfo{person}{Yannic
  Noller}, \bibinfo{person}{Lars Grunske}, {and} \bibinfo{person}{Abhik
  Roychoudhury}.} \bibinfo{year}{2021}\natexlab{}.
\newblock \showarticletitle{Concolic Program Repair}. In
  \bibinfo{booktitle}{\emph{Proceedings of the 42nd ACM SIGPLAN International
  Conference on Programming Language Design and Implementation}} (Virtual,
  Canada) \emph{(\bibinfo{series}{PLDI 2021})}. \bibinfo{publisher}{Association
  for Computing Machinery}, \bibinfo{address}{New York, NY, USA},
  \bibinfo{pages}{390–405}.
\newblock
\showISBNx{9781450383912}
\urldef\tempurl%
\url{https://doi.org/10.1145/3453483.3454051}
\showDOI{\tempurl}


\bibitem[Smith et~al\mbox{.}(2015)]%
        {Smith15}
\bibfield{author}{\bibinfo{person}{Edward~K. Smith}, \bibinfo{person}{Earl~T.
  Barr}, \bibinfo{person}{Claire Le~Goues}, {and} \bibinfo{person}{Yuriy
  Brun}.} \bibinfo{year}{2015}\natexlab{}.
\newblock \showarticletitle{Is the Cure Worse than the Disease? Overfitting in
  Automated Program Repair}. In \bibinfo{booktitle}{\emph{Proceedings of the
  2015 10th Joint Meeting on Foundations of Software Engineering}} (Bergamo,
  Italy) \emph{(\bibinfo{series}{ESEC/FSE 2015})}.
  \bibinfo{publisher}{Association for Computing Machinery},
  \bibinfo{address}{New York, NY, USA}, \bibinfo{pages}{532–543}.
\newblock
\showISBNx{9781450336758}
\urldef\tempurl%
\url{https://doi.org/10.1145/2786805.2786825}
\showDOI{\tempurl}


\bibitem[Soremekun et~al\mbox{.}(2022)]%
        {Soremekun22}
\bibfield{author}{\bibinfo{person}{Ezekiel Soremekun}, \bibinfo{person}{Esteban
  Pavese}, \bibinfo{person}{Nikolas Havrikov}, \bibinfo{person}{Lars Grunske},
  {and} \bibinfo{person}{Andreas Zeller}.} \bibinfo{year}{2022}\natexlab{}.
\newblock \showarticletitle{Inputs From Hell:}.
\newblock \bibinfo{journal}{\emph{IEEE Transactions on Software Engineering}}
  \bibinfo{volume}{48}, \bibinfo{number}{4} (\bibinfo{year}{2022}),
  \bibinfo{pages}{1138--1153}.
\newblock
\urldef\tempurl%
\url{https://doi.org/10.1109/TSE.2020.3013716}
\showDOI{\tempurl}


\bibitem[Sun et~al\mbox{.}(2016)]%
        {Sun16}
\bibfield{author}{\bibinfo{person}{Chengnian Sun}, \bibinfo{person}{Vu Le},
  {and} \bibinfo{person}{Zhendong Su}.} \bibinfo{year}{2016}\natexlab{}.
\newblock \showarticletitle{Finding Compiler Bugs via Live Code Mutation}.
\newblock \bibinfo{journal}{\emph{SIGPLAN Not.}} \bibinfo{volume}{51},
  \bibinfo{number}{10} (\bibinfo{date}{oct} \bibinfo{year}{2016}),
  \bibinfo{pages}{849–863}.
\newblock
\showISSN{0362-1340}
\urldef\tempurl%
\url{https://doi.org/10.1145/3022671.2984038}
\showDOI{\tempurl}


\bibitem[{van Tonder} and {Le Goues}(2018)]%
        {TonderG18}
\bibfield{author}{\bibinfo{person}{Rijnard {van Tonder}} {and}
  \bibinfo{person}{Claire {Le Goues}}.} \bibinfo{year}{2018}\natexlab{}.
\newblock \showarticletitle{Static Automated Program Repair for Heap
  Properties}. In \bibinfo{booktitle}{\emph{ICSE}}. \bibinfo{publisher}{{ACM}},
  \bibinfo{pages}{151--162}.
\newblock


\bibitem[Wang et~al\mbox{.}(2017)]%
        {Bo17}
\bibfield{author}{\bibinfo{person}{Bo Wang}, \bibinfo{person}{Yingfei Xiong},
  \bibinfo{person}{Yangqingwei Shi}, \bibinfo{person}{Lu Zhang}, {and}
  \bibinfo{person}{Dan Hao}.} \bibinfo{year}{2017}\natexlab{}.
\newblock \showarticletitle{Faster Mutation Analysis via Equivalence modulo
  States}. In \bibinfo{booktitle}{\emph{Proceedings of the 26th ACM SIGSOFT
  International Symposium on Software Testing and Analysis}} (Santa Barbara,
  CA, USA) \emph{(\bibinfo{series}{ISSTA 2017})}.
  \bibinfo{publisher}{Association for Computing Machinery},
  \bibinfo{address}{New York, NY, USA}, \bibinfo{pages}{295–306}.
\newblock
\showISBNx{9781450350761}
\urldef\tempurl%
\url{https://doi.org/10.1145/3092703.3092714}
\showDOI{\tempurl}


\bibitem[Yan et~al\mbox{.}(2016)]%
        {Hua16}
\bibfield{author}{\bibinfo{person}{Hua Yan}, \bibinfo{person}{Yulei Sui},
  \bibinfo{person}{Shiping Chen}, {and} \bibinfo{person}{Jingling Xue}.}
  \bibinfo{year}{2016}\natexlab{}.
\newblock \showarticletitle{Automated Memory Leak Fixing on Value-Flow Slices
  for C Programs}. In \bibinfo{booktitle}{\emph{Proceedings of the 31st Annual
  ACM Symposium on Applied Computing}} (Pisa, Italy)
  \emph{(\bibinfo{series}{SAC '16})}. \bibinfo{publisher}{Association for
  Computing Machinery}, \bibinfo{address}{New York, NY, USA},
  \bibinfo{pages}{1386–1393}.
\newblock
\showISBNx{9781450337397}
\urldef\tempurl%
\url{https://doi.org/10.1145/2851613.2851773}
\showDOI{\tempurl}


\end{thebibliography}

\end{document}